\pdfoutput=1
\documentclass[%
prl,twocolumn,
notitlepage,
 amsmath,amssymb,
 aps,
floatfix,
longbibliography
]{revtex4-1}
\usepackage[utf8]{inputenc}
\usepackage[T1]{fontenc}

\usepackage{color}
\usepackage{amsmath}
\usepackage{amssymb}
\usepackage{bm}
\usepackage{graphicx}
\usepackage{xcolor}
\usepackage[%
  colorlinks=true,
  urlcolor=blue,
  linkcolor=blue,
  citecolor=blue
]{hyperref}
\usepackage{url}
\usepackage{array}
\usepackage{siunitx}

\newcommand{\ind}[1]{_{\mathrm{#1}}}
\renewcommand\vec{\mathbf}
\newcommand{\bb}{\vec{b}}

\newcommand{\ee}{\vec{e}}

\newcommand{\FF}{\vec{F}}

\newcommand{\rr}{\vec{r}}
\newcommand{\RR}{\vec{R}}
\newcommand{\uu}{\vec{u}}

\newcommand{\pp}{\vec{p}}
\newcommand{\qq}{\vec{q}}

\newcommand{\VV}{\vec{V}}

\begin{document}
\title{When soft crystals defy Newton's third law: \texorpdfstring{\\}{} Non-reciprocal mechanics and  dislocation motility }
\author{Alexis Poncet}
\email{alexis.poncet@ens-lyon.fr}
\affiliation{ Univ. Lyon, ENS de Lyon, Univ. Claude Bernard, CNRS, Laboratoire de Physique, F-69342, Lyon.}
\author{Denis Bartolo}
\email{denis.bartolo@ens-lyon.fr}
\affiliation{ Univ. Lyon, ENS de Lyon, Univ. Claude Bernard, CNRS, Laboratoire de Physique, F-69342, Lyon.}

\begin{abstract}
The  effective interactions between the constituents of driven soft matter generically defy Newton's third law. Combining theory and numerical simulations, we establish that six
classes of mechanics with no counterparts in equilibrium systems emerge in elastic crystals challenged by nonreciprocal interactions. Going beyond linear deformations, we reveal that interactions violating Newton's third law generically turn otherwise quiescent dislocations into motile singularities which steadily glide though periodic lattices.
\end{abstract}
\date{\today}
\maketitle

\begin{figure*}
    \includegraphics[scale=1]{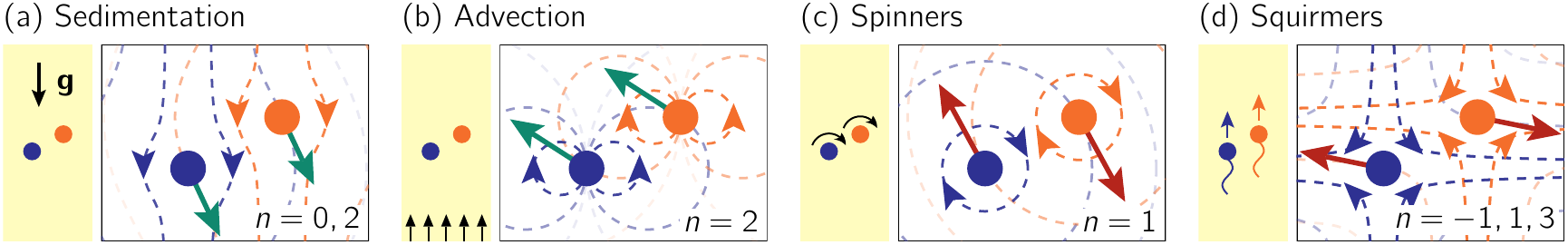}
    \caption{{\bf Hydrodynamic interactions do not obey Newton's third law.}  We give four examples of non-reciprocal forces between pairs of identical particles driven in a viscous fluid. Parity-symmetric forces (green arrows) % result in
    correspond to equal drag forces acting on the particles.
    {\bf (a)} Two identical particles sedimenting in a viscous fluid.  The interaction force field decomposes on the $n=0,2$ modes of Eq.~\eqref{Eq:angular}.
    {\bf (b)} Two particles advected in a shallow microfluidic channel interact hydrodynamically via the sole $n=2$ mode.
    Parity-antisymmetric forces (red arrows) 
    correspond to opposite drag forces acting on  the particles.
    {\bf (c)} Two particles spinning in a viscous fluid interact via transverse forces  ($n=1$ mode only).
    {\bf (d)} When swimming in the same direction two so-called squirmers interact hydrodynamically via the superposition of forces with $n=-1,\,1\,\rm and\, 3$ angular modes (see SI).}
    \label{fig:1}
\end{figure*}

The  constituents of driven soft matter interact via  effective forces which   are generically non-reciprocal. These constituents escape the elementary constraints imposed by Newton's third law by constantly  exchanging  linear and angular momentum with their surrounding medium.
Hydrodynamic interactions provide a paradigmatic example of such non-reciprocal interactions~\cite{Happel_book}. 
Take the minimal system involving two identical colloidal particles $A$ and  $B$ sedimenting in a viscous fluid, the drag force acting on $B$ due to the motion of $A$ is never opposite to the drag force acting on $A$ due to the motion of $B$, whatever the relative position of the two particles, see Fig.~\ref{fig:1}a. 
In a colloidal suspension these  non-reciprocal interactions result in complex chaotic trajectories at odds with the homogeneous and steady nature of the global drive~\cite{Guazzelli_Review,Ramaswamy_2001}. As illustrated in Fig.~\ref{fig:1}a-d,  the violation of Newton's third law is not specific to sedimentation but applies to systems as diverse as colloidal spinners~\cite{Yan2015,Petroff2015,Soni2019,Bililign_2021}, driven emulsions~\cite{Beatus2006,Beatus2007,Desreumaux_2012} and swimmer suspensions~\cite{Lauga2020}.
Beyond the specifics of fluid mechanics, non-reciprocal couplings rule
systems as diverse as self-phoretic colloids~\cite{Soto_2014,Saha2014,Golestanian2019}, programmable  matter~\cite{Lavergne2019,Fruchart_2020}, dirty plasmas~\cite{Ivlev_2015}, groups of living creatures~\cite{Cavagna_Review,Moussaid2011,Eloy2018} and %, or
motile agents~\cite{Vicsek95,Chate2017,Dadhichi_2020}. 
Despite a surge of recent efforts, see e.g. Refs~\cite{Coulais2017,Brandenbourger2019,Golestanian2020,Scheibner_2020,Fruchart_2020}, the basic principles relating the violation of Newton's third law at the microscopic level  to the phase behavior, and mechanics of driven soft matter remains elusive and limited to specific realizations.

In this letter, we investigate the
mechanics of crystals assembled from (self-)driven units defying
Newton's third law. We  lay out  a comprehensive description of their elastic responses, by
classifying  non-reciprocal microscopic  interactions in term of their  symmetry under parity transformation. Building on this framework, we then combine theory and simulations to show that driven lattices generically feature a macroscopic response that defies our intuition based on solid mechanics, and host self-propelled dislocations.

Let us consider the overdamped dynamics of a collection of $N$ identical  point particles in a homogeneous medium. The particles, located at positions $\RR^\nu(t)$, are supposed to interact via pairwise-interaction forces balanced by a local frictional drag $-\zeta \dot{\RR}^\nu(t) $. Their equations of motion  reduce to the generic form~\footnote{The dynamics also reduces to Eq.~(1) when a constant driving force $\mathbf F_0$ acts on all particles (by setting $\RR\to\RR-\FF_0t$).}
\begin{equation}
    \zeta \dot\RR^\nu =  \sum_{\mu\neq \nu} \FF(\RR^\mu-\RR^\nu),
    \label{Eq:Eq1}
\end{equation}
and we henceforth set $\zeta=1$ without loss of generality.
Having ignored the many degrees of freedom of the medium hosting the $N$ particles, we stress that the effective force $\FF$ is not constrained to derive from  any interaction potential (or free energy). 
It includes all effective couplings mediated by the soft matrix surrounding the particles. 
For instance, $\FF$ can reflect the hydrodynamic, or phoretic, interactions experienced by  particles driven in a viscous fluid~\cite{Happel_book,Saha2014,Lauga2020}, or the elastic forces acting on active inclusions deforming an elastic solid~\cite{Ramaswamy2000,Kumar2014,Ramananarivo2019,Gupta_2020}.
Beyond the specifics of these  examples, the $N$ particles can continuously exchange linear and angular momentum with their surrounding environment, thereby alleviating the constraints imposed by Newton's third law. In general we have (i) $\FF(\RR^\mu-\RR^\nu)\neq-\FF(\RR^\nu-\RR^\mu)$ and (ii) $\RR\times \FF(\RR)\neq\vec{0}$. We henceforth refer to forces obeying (i)
or (ii) as non-reciprocal forces. Our first goal is to show that non-reciprocal forces cannot stabilize crystalline order on their own.

We first note  that all  periodic lattices  correspond to a stationary solution of Eq.~\eqref{Eq:Eq1}.
To single out the consequences of force nonreciprocity, we decompose  $\FF$ on its symmetric and antisymmetric components under parity transformation: $\FF(\RR)=\FF_{\rm A}(\RR)+\FF_{\rm S}(\RR)$, where  {the parity-symmetric interactions} ($\FF_{\rm S}(-\RR)=\FF_{\rm S}(\RR)$) contribute to linear momentum exchange with the surrounding medium, while
the parity-antisymmetric interactions
($\FF_{\rm A}(-\RR)=-\FF_{\rm A}(\RR)$)  reflect orbital momentum exchange. To gain some intuition, we show two prototypical examples of parity-symmetric interactions in Fig.~\ref{fig:1}a-b, and contrast them with two classical examples of antisymmetric hydrodynamic interactions in  Fig.~\ref{fig:1}c-d.

In crystals,
translational invariance then implies that $\FF_{\rm A}$ does not contribute to Eq.~\eqref{Eq:Eq1}, and that $\FF_{\rm S}$ can merely power the steady translation of periodic lattices, without altering their structure. Parity-symmetric interactions uniformly translate the particles at a constant speed given by $\VV_0=\sum_{\mu\neq0} \FF_{\rm S}(\RR^\mu_0)$, where the $\RR^\mu_0$ are the positions in the crystal frame centered on $\RR_0^0=\bm 0$. 
To address the stability of  driven lattices, we now consider small amplitude displacements taking the form of plane waves  $\uu(\qq, t) e^{-i\qq\cdot\RR^\nu_0}$ (in the crystal frame). 
The structural stability is then determined by the eigenvalues of the linear dynamics given by
\begin{align}
    \dot\uu(\qq, t) &= (M\ind{A} + i M\ind{S})\cdot\uu(\qq, t)
    \label{Eq:Stability}
\end{align}
where the stability matrices are both real and read 
\begin{align}
    M\ind{A} &= -\sum_{\mu\neq 0}\left[1-\cos(\qq\cdot\RR^\mu_0)\right]\bm\nabla\FF\ind{A}(\RR^\mu_0), 
    \label{Eq:MA}
    \\
    M\ind{S} &= -\sum_{\mu\neq 0}  \sin(\qq\cdot\RR^\mu_0)\bm\nabla\FF\ind{S}(\RR^\mu_0).
    \label{Eq:MS}
\end{align}    
$\bm \nabla\FF$ is  the Jacobian matrix associated with $\FF(\RR)$, see SI.  Eq.~\eqref{Eq:Stability} reduces to the trivial stability of overdamped phonons in passive crystals only when $\FF\ind{A}$ is  longitudinal and $\FF\ind{S}=\vec{0}$.  
Addressing separately the role of parity-symmetric and antisymmetric forces, we show how
non-reciprocal forces either sustain the propagation of deformation waves in overdamped systems, or compete with potential  interactions to destabilize crystalline order. 
We focus on 2D crystals and ignore the possible interplay between the local drag and structural deformations~\cite{Lahiri97,Chajwa2020}. 
Under the action of $\FF\ind{S}$   the eigenvalues of $M_S$ (Eq.~\eqref{Eq:MS}) take the form $\lambda_\pm=\frac{1}{2}\left\{{\rm tr}(M\ind{S})\pm\left[{\rm tr}(M\ind{S})^2-4{\det}(M\ind{S})\right]^{1/2}\right\}$. It then readily follows that $i\lambda_+$ and $i\lambda_-$ cannot be both real negative numbers thereby prohibiting the existence of overdamped phonons. 
 We note that this prediction correctly  accounts for the free propagation of phonons powered by parity-symmetric hydrodynamic interactions in driven microfluidic crystals~\cite{Beatus2006,Beatus2007,Desreumaux_2012}.
The effect of parity-antisymmetric forces is more subtle. However in a number of practical realizations, $\FF\ind{A}$ is  divergenceless. This is the case of hydrodynamic interactions  mediated by an incompressible fluid, of phoretic interactions  induced by a rapidly diffusing solute~\cite{Soto_2014,Golestanian2019} and of all mechanisms inducing  transverse couplings as in the odd-elastic crystals discussed in Refs.~\cite{Scheibner_2020,Bililign_2021,Tan2021}. When the condition $\bm \nabla\cdot \FF\ind{A}=0$ is met, ${\rm tr}(M_{\rm A})=0$, and therefore  the eigenvalues of $M_{\rm A}$ are either opposite real or pure imaginary numbers. Under the sole action of divergenceless antisymmetric forces driven crystals  therefore either support free phonons or are linearly unstable in agreement with the phenomenology reported in colloidal spinner crystals~\cite{Bililign_2021,Tan2021}. 
In sum, regardless of their microscopic origin, we have shown that non-reciprocal forces are typically unable to self-organize identical units into a stable crystal state.

\begin{figure*}
    \includegraphics[scale=1]{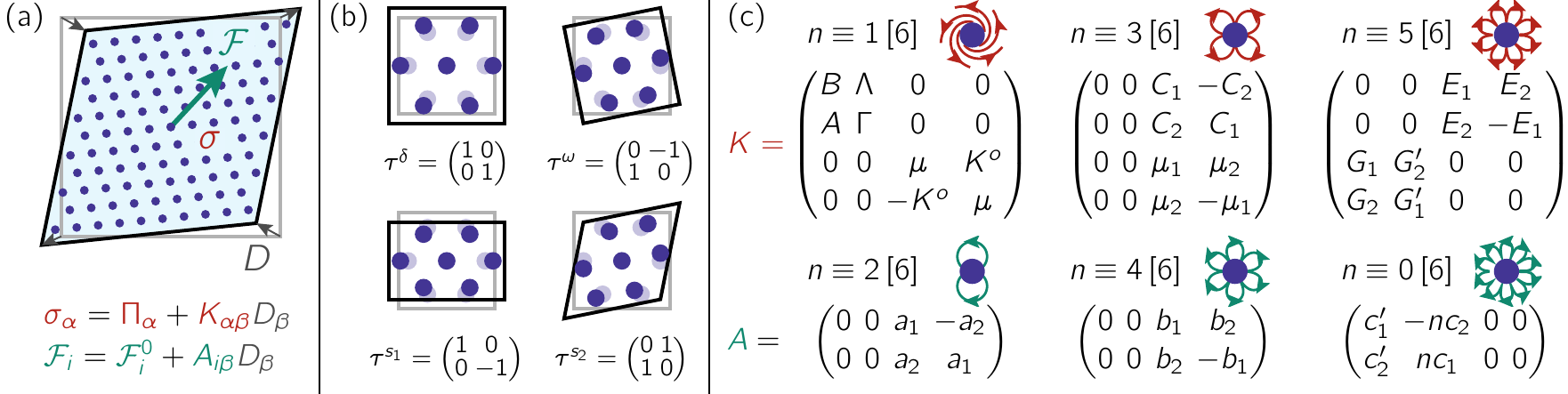}
    \caption{{\bf Mechanics of elastic lattices challenged by nonreciprocal interactions.}
    {\bf (a)} When a hexagonal lattice undergoes a deformation, parity-antisymmetric forces %$\FF_A$
    induce a stress $\sigma$ (red) that can be computed using the Irving-Kirkwood formula (Eq.~\eqref{eq:IrvingKirkwood}). The macroscopic response to a deformation $D$ reads $\sigma_\alpha = \Pi_\alpha + K_{\alpha\beta} D_\beta$ where  $\alpha, \beta$ denote the elementary stress and deformation modes  defined in {\bf (b)} and $\Pi$ is a so-called prestress supported by the crystal even in absence of any deformations.
    In the case of parity-symmetric interactions, % $\FF_s$,
    a deformation $D$ induces a net force $\mathcal{F}$ defined as the net force experienced by any test particle (green arrow). $\mathcal{F}$ and $D$ are related by the linear yet noninvertible relation $\mathcal{F}_i = \mathcal{F}_i^0 + A_{i\beta} D_\beta$ where $i=x, y$. The six possible structures for the matrices $K$ and $A$ are given in {\bf (c)}.
    {\bf (b)} Elementary modes of 2D deformations (and stresses).
    $\tau^\delta$ corresponds to a dilation (isotropic pressure),
    $\tau^\omega$ to a rotation (torque)
    while $\tau^{s_1}$ and $\tau^{s_2}$ are the two pure-shear deformations (shear stresses).
    {\bf (c)} On a hexagonal lattice, parity-antisymmetric interactions define three classes of stress-strain relations. These constitutive relations are determined by the angular symmetry of the interactions ($n\equiv 1, 3, 5\, [6]$ where `$[6]$' stands for `modulo $6$'), see Eq.~\eqref{Eq:angular}. The prestresses $\Pi$ are given:
    $\Pi = (P_\delta, T_\omega, 0, 0)$ when $n\equiv 1\, [6]$, $\Pi = 0$ when $n\equiv 3\, [6]$,  and $\Pi = (0, 0, S_1, S_2)$ when $n\equiv 5\, [6]$.
    Parity-symmetric interactions define three classes of force-strain relations determined by the angular symmetry of the interactions ($n\equiv 2, 4, 0\, [6]$). The three possible constitutive relations correspond to the three possible structures of the matrix $A$. A net force $\mathcal{F}_i^0$  can exist even in the absence of deformation under the action of nonreciprocal interactions when $n\equiv 0\, [6]$, it is given by $\mathcal{F}_i^0=(c_1, c_2) \neq 0$.
    The analytic expression of all material parameters are expressed in term of the microscopic interactions both for  parity-antisymmetric and symmetric forces in the SI.
    }
    \label{fig:2}
\end{figure*}

To understand how non-reciprocal forces modify the mechanics of elastic crystals, it is useful to classify the interactions with respect to their angular symmetries. In all that follows we consider a generic class of forces  which we decompose in a multipolar expansion as
\begin{equation} 
    \FF(\RR = re^{i\theta}) = \sum_{n} f_n(r) e^{i(n\theta - \alpha_n)},
    \label{Eq:angular}
\end{equation}
where we use the complex-number representation for the 2D vectors.
Forces are parity-symmetric when $n$ is even, and antisymmetric when $n$ is odd. 
Given this simple decomposition Newton's third law would reduce the sum in Eq.~\eqref{Eq:angular} to its sole $n=1$ mode, and imply $\alpha_1=0$.
    At lowest orders, non-reciprocal interactions correspond to forces with $n=1,\,\alpha_1=\pi/2$ and $n=2$. 
They are realized by two paradigmatic examples of driven soft matter.
The  parity-odd case $n=1,\,\alpha_1=\pi/2$, corresponds to the forces experienced by  collections of active or driven spinners interacting either by near-, or far-field hydrodynamic interactions in a viscous fluid, see Fig.~\ref{fig:1}c and e.g. Refs.~\cite{Yan2015,yeo2015,Petroff2015,Soni2019,Oppenheimer2019,Shen2020}. 
Similarly, the parity-even case $n=2$ corresponds to the standard dipolar flows ruling the interactions between foams, emulsions, or colloids  uniformly driven in  shallow channels, see see Fig.~\ref{fig:1}b and e.g. Refs.~\cite{Beatus2006,Desreumaux_2013,Shani2014,Mondal2020}. 

We focus here on perfect hexagonal lattices, and compute the stresses, $\sigma$, and body forces, $\mathcal F$, resulting from linear deformations. In line with our intuition based on equilibrium  solids, we show below that parity-odd interactions convert deformations into stresses. However, parity-even forces do not cancel one another. As a result, deformations generically produce net forces. 

Without loss of generality we define the real space displacement field  $\mathbf u(\rr)$, and decompose the deformations tensor $D_{ij}=\partial u_i/\partial x_j$,  on the basis defined by the pure dilation ($\tau^\delta_{ij}$), rotation ($\tau^\omega_{ij}$), and two orthogonal shear modes ($\tau^{s_1}_{ij}$, and $\tau^{s_2}_{ij}$), see Fig.~\ref{fig:2}b, %~\cite{Scheibner_2020}:
\begin{equation} \label{Eq:D}
    D_{ij} = D_\delta \tau^\delta_{ij} + D_\omega \tau^\omega_{ij}
    + D_{s_1} \tau^{s_1}_{ij} +  D_{s_2} \tau^{s_2}_{ij}.
\end{equation}
%See also SI for details and illustrations.

{\em Parity even interactions.} To illustrate the counter intuitive mechanics induced by parity even forces, we first focus  on the minimal example of interactions having a dipolar symmetry ($n=2$, see Fig.~\ref{fig:1}b). Computing the force $\mathcal F$ acting on a test particle in a deformed hexagonal lattice,  %Fig~\ref{fig:2}a,
we find that the force components ($\mathcal F_i$) depend linearly on deformation ($D_{ij}$):
\begin{equation} \label{Eq:force_dipole}
    \mathcal{F}
    = K_2 \begin{pmatrix}  \cos\alpha_2 & \sin\alpha_2 \\ -\sin\alpha_2 &  \cos\alpha_2 \end{pmatrix}
    \begin{pmatrix}
     D_{s_1} \\ D_{s_2}
    \end{pmatrix},
\end{equation}
where $K_2=\sum_{\mu\neq0} \left[f_2(R^\mu_0) + R^\mu_0 f_2'(R^\mu_0)/2\right]$. 
The consequences of Eq.~\eqref{Eq:force_dipole} are clear. 
Having in mind a colloidal crystal driven in a 2D fluid, say in the $x$ direction (leading to $\alpha_2=0$), longitudinal shear deformations ($D_{s_1}$) accelerate or slow down  translational motion whereas shearing the lattice at a $45^\circ$ angle ($D_{s_2}$) results in a net drift in the direction transverse to the applied drive. Rotation and dilation do not yield any net force. 
We emphasize that this atypical  response relating strain to body forces is  linear but not invertible. It therefore violates any form of macroscopic reciprocity~\cite{Coulais2017,Fruchart_2020}. Uniform deformations cause net uniform forces, but applying a homogeneous force merely  translates the crystal leaving its inner structure unaltered. The emergence of net body forces from crystal deformations is not specific to dipolar interactions, but generically emerges from non-reciprocal forces even under parity transformations. In Fig.~\ref{fig:2}c, we show the only three possible constitutive relations in crystals enjoying a six-fold symmetry defined by the response matrix $A$: $\mathcal F_i=\mathcal F_i^0 +A_{i\beta}D_\beta$ where the implicit summation ($\beta$ index) is done over the four deformation modes.

\begin{figure*}
    \centering
    \includegraphics{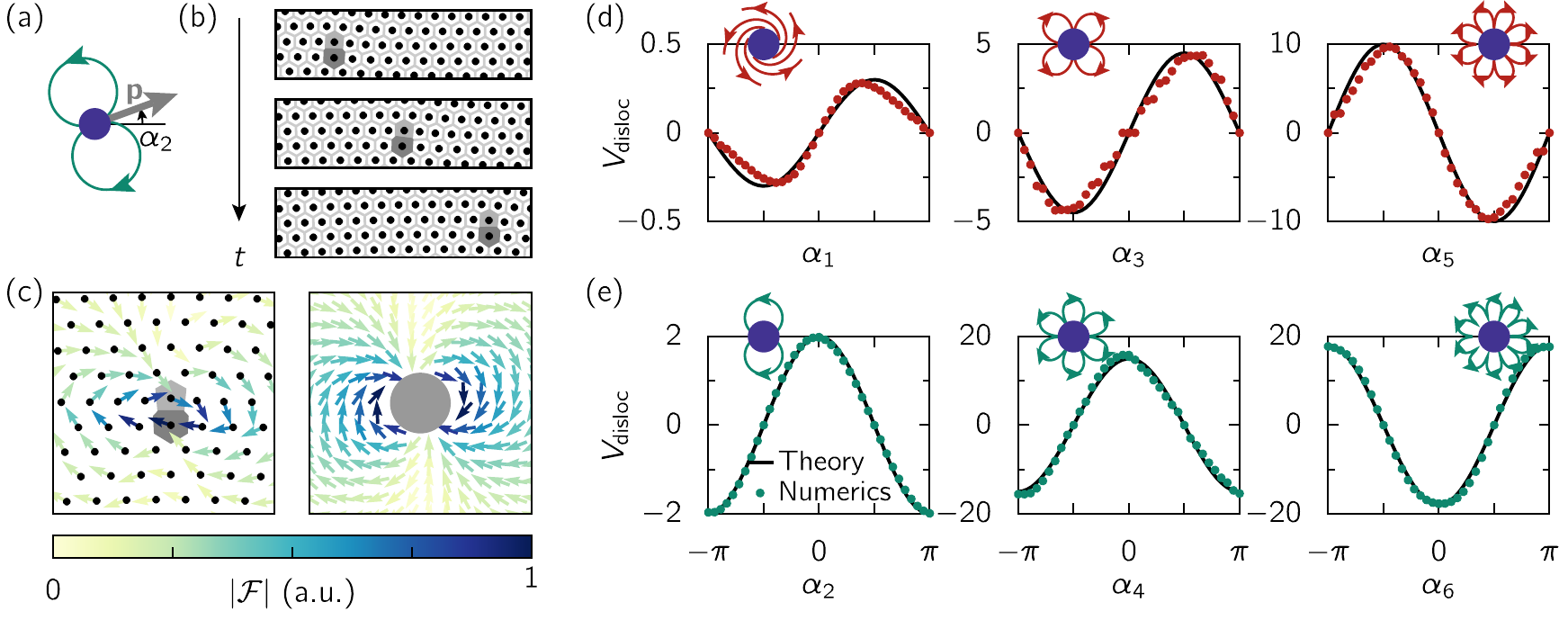}
    \caption{{\bf Non-reciprocal interactions powers dislocation  glide}.
    In our numerical simulations (detailed in SI), we choose the non-reciprocal interactions to correspond to pure multipoles. When $n\geq 2$, $f_n(r) = 1/r^n$ (see Eq.~\eqref{Eq:angular}). The case $n=2$ then corresponds to hydrodynamic interactions in shallow channels. For $n=1$, we arbitrarily choose $f_1(r) = e^{-r^2/2}$ to reflect the short range coupling between colloidal spinners. Furthermore, the crystal is stabilized by reciprocal interactions $\FF(\rr) = -A\ind{rep}\rr/\|\rr\|^5$ (dipole-dipole repulsion).
    {\bf (a)} Sketch of dipolar field ($n=2$) induced by a given particle. $\alpha_2$ is the phase of mode $n=2$ and $\pp = (\cos\alpha_2, \sin\alpha_2)$.
    {\bf (b)} Glide of a dislocation in a crystal. The non-reciprocal interactions correspond to $n=2$, $\alpha_2 = 20\si{\degree}$ ($A\ind{rep} = 10$). The three snapshots correspond to $t = 0,\,0.25$ and $0.5$.
    {\bf (c)} Left: numerical computation of the forces on the particles induced by dipolar pairwise interactions ($\alpha_2= 20\si{\degree}$). Right: theoretical prediction for the force field according to Eq~\eqref{Eq:force_dip}, for the same microscopic parameters.
    {\bf (d)} The variations of the gliding speed of a dislocation with the phase $\alpha_n$ measured from our numerical simulations (red dots) are in excellent agreement with our theoretical predictions (black solid line). From left to right $n=1, 3$ and $5$ ($A\ind{rep} = 1, 10, 50$). 
    {\bf(e)} Same plots as in {(\bf d)} in the case of parity-even forces. From left to right $n=2, 4$ and $6$ ($A\ind{rep} = 10, 20, 50$).
    }
    \label{fig:3}
\end{figure*} 

{\em Parity-odd forces.} The case of parity-odd forces is more familiar. Let us start again with a simple example, when the microscopic forces are isotropic ($n=1,\,\alpha_1\neq0$, see Fig.~\ref{fig:1}c), our hexagonal lattice realizes a typical example of an  odd-elastic solid introduced in Ref.~\cite{Scheibner_2020} and exemplified in Ref.~\cite{Bililign_2021}.
Deformations do not yield body force but build up a net stress which we can readily compute using the  Irving-Kirkwood formula~\cite{Irving_1950,Yang2012,Klymko2017},
\begin{equation} \label{eq:IrvingKirkwood}
    \sigma_{ij}=\frac{1}{V}\sum_{\mu\neq 0} R_i^\nu F_j(\RR^\mu),
\end{equation}
where the summation excludes the particle at the origin and $V$ is the volume of the unit cell. 
Decomposing the stress on the dilation, rotation, and two shear modes $(\sigma_\delta, \sigma_\omega, \sigma_{s_1}, \sigma_{s_2})$ we then find the constitutive relation
\begin{equation} \label{Eq:oddelasticity}
    \begin{pmatrix}
    \sigma_\delta \\ \sigma_\omega \\ \sigma_{s_1} \\ \sigma_{s_2}
    \end{pmatrix}
    = \begin{pmatrix}
    P_\delta \\ T_\omega \\ 0 \\ 0
    \end{pmatrix}
    + \begin{pmatrix} B & 0 & 0 & 0 \\ A & 0 & 0 & 0 \\
    0 & 0 & \mu & K^o \\ 0 & 0 & -K^o & \mu \end{pmatrix}
    \begin{pmatrix}
    D_\delta \\ D_\omega \\ D_{s_1} \\ D_{s_2}
    \end{pmatrix},
\end{equation}
where the analytic expressions of all material parameters are presented in SI together with an alternative geometrical derivation of $\sigma$. 
Two comments are in order. First, the undeformed crystal can support both a non-zero pressure,  $P_{\delta}$, and a non-zero odd stress, $T_{\omega}$ viz. a net torque density resulting from the angular momentum exchange with the surrounding medium. This result confirms early phenomenological theories, simulations and recent experiments conducted on colloidal and granular spinners~\cite{Dahler1961,Tsai2005,VanZuiden2016,Soni2019}. 
Secondly, we find that  
the bulk  and shear moduli ($B$ and $\mu$) both vanish for purely transverse forces ($\alpha_1=\pi/2$), 
whereas odd elasticity (coefficients $A$ and $K^o$) emerges when $\alpha_1\neq0$.
Remarkably, in hexagonal lattices all microscopic interactions satisfying $n\equiv 1\ (\mathrm{mod}\ 6)$ yield 
the
same macroscopic elastic response, which expands the existence of odd elastic solids beyond the specifics of purely transverse forces.
More generally, we summarize our results obtained for all odd $n$ in Fig.~\ref{fig:2}c, and show the only three  possible affine constitutive relations of non-reciprocal hexagonal crystals.
When $|n|\geq3$ elasticity is not isotropic anymore, even in hexagonal lattices.
We also note that when  $n \equiv 5\ (\mathrm{mod}\ 6)$ undeformed lattices support a net shear stress {$(\sigma_{s_1}, \sigma_{s_2}) = (S_1, S_2)$} mirroring the odd stress of chiral crystals~\cite{Tsai2005,VanZuiden2016,Soni2019}.

Our continuum mechanics picture provides an effective platform to go beyond linear deformations, and address the impact of  dislocations on the crystal structure and dynamics. 
To gain some  insight, we first solve Eq.~\eqref{Eq:Eq1} numerically using periodic boundary conditions, for a hexagonal lattice deformed by two  maximally separated  dislocations of opposite Burgers vector (periodic boundary conditions require a zero net topological charge).  
The numerical methods are detailed in SI. 
We investigate separately the impact of each angular mode. 
Remarkably, Supplementary video 1 reveals that the competition between 
non-reciprocal forces and  elasticity  turns dislocations into self-propelled singularities.
Fig.~\ref{fig:3}b  illustrates the gliding motion of dislocations powered by dipolar ($n=2$) interactions.  

To explain this spontaneous  motion in the direction of the Burgers vector $\mathbf b$,  we consider an isolated dislocation at the origin
of an isotropic elastic solid  ($B\neq0$ and $\mu\neq0$ in Eq.~\eqref{Eq:oddelasticity}). In the  case of parity-odd forces, the induced  internal extra stress $\Pi$, see Fig.~\ref{fig:2}, results in a net  Peach-Koehler force $F^{PK}_i=\epsilon_{ij}\Pi_{kj}b_k$, computed in SI.  When $\bb=b\ee_1$, the glide component of the force takes a compact form when $n\equiv 1, 5\ (\mathrm{mod}\ 6)$:
\begin{align}
F^{PK}_{\rm glide}(n=1)&=-T_\omega b & F^{PK}_{\rm glide}(n=5)=S_2 b.
\end{align}
The case $n=3$ deserves a separate study reported in SI, which accounts for the local rotation of the crystal orientation. 
Beyond the existence of a nonvanishing Peach-Koehler force, our theoretical predictions quantitatively account for the variations of the dislocation speed as a function of the microscopic force's phase $\alpha_n$ shown in Fig.~\ref{fig:3}d. Our finding confirms and generalize the experimental and numerical results reported in odd colloidal crystals ($n=1$)~\cite{Bililign_2021}.

The influence of parity-even forces, is yet again more subtle as they do not translate in macroscopic stresses but net forces. This essential difference requires expanding the conventional Peach-Koehler picture. We can  gain some intuition by looking at the symmetry of  the microscopic-force distribution  around an isolated dislocation when perturbed by dipolar forces ($n=2$), see Fig.~\ref{fig:3}c. It clearly reveals a simple shear component, which is  correctly captured by our continuum theory (Eq.~\eqref{Eq:force_dipole}), see also SI. Using the strain field around an isolated dislocation~\cite{Landau,Braverman_2020}, the force field takes the  form    
\begin{equation} \label{Eq:force_dip}
    \mathcal{F}(r, \theta) = {\mathcal F_0} \frac{\bb\cdot\rr}{r^2}
    \begin{pmatrix}
    \sin(2\theta-\alpha_2) \\ -\cos(2\theta-\alpha_2)
    \end{pmatrix},
\end{equation}
where $\mathcal F_0=-{[K_2 (1+\nu)]}/{4\pi }$, and $\nu=(B-\mu)/(B+\mu)$. 
We can then define an effective Peach-Koehler force acting on the dislocation core as $F^{\rm PK}_i=\epsilon _{ij}\partial_k{\mathcal F}_j b_k$. To evaluate it, we  need to regularize the force gradient at $r=0$. As detailed in SI, we evaluate $\FF^{\rm PK}$ as a weighted integral over a circle of radius $b$ around the dislocation core. We finally find:
\begin{equation}
\FF^{PK} = \frac{1}{4} \mathcal{F}_0\pp,
\end{equation}
where $\mathbf p=(\cos\alpha_2,\sin\alpha_2)$ is the vector defining the dipole orientation (Fig.~\ref{fig:3}a).  This expression indicates that  non-reciprocal dipolar forces can sustain the gliding motion  of dislocations having a Burgers vector making a finite angle with  $\pp$. To further confirm this  prediction, we use our numerical simulations and plot the glide velocity as a function of the angle $\alpha_2$. Fig.~\ref{fig:3}e shows a remarkable agreement with our effective elastic theory predicting a speed proportional to $\cos \alpha_2$. As a last result, we stress  that beyond the specifics of dipolar forces,  the emergent Peach-Koehler forces evaluated at the continuum level correctly accounts for the variations of the dislocation speed measured in our numerical simulations whatever the angular symmetry of the non-reciprocal forces (Fig.~\ref{fig:3}e).

We have elucidated the relations between the microscopic symmetries of non-equilibrium interactions and the macroscopic response of driven lattices.  From a practical perspective, the systematic classification summarized in Fig.~\ref{fig:2} suggests effective strategies to design active metamaterials having mechanical properties out of reach of equilibrium systems.  
From a fundamental perspective, our findings immediately raise three basic questions: how does dislocation motility redefine the plastic flows, fracture and melting of driven and active crystals?

\begin{acknowledgments}
We thank  E. Billilign, Y. Ganan, W. T. M Irvine, and M. Le Blay and V. Vitelli for invaluable discussions and suggestions. This work was partly supported by Idex ToRe and ANR grant WTF.
\end{acknowledgments}

\end{document}

\title{When soft crystals defy Newton's third law: \texorpdfstring{\\}{} Non-reciprocal mechanics and  dislocation motility }
\author{Alexis Poncet}
\email{alexis.poncet@ens-lyon.fr}
\affiliation{ Univ. Lyon, ENS de Lyon, Univ. Claude Bernard, CNRS, Laboratoire de Physique, F-69342, Lyon.}
\author{Denis Bartolo}
\email{denis.bartolo@ens-lyon.fr}
\affiliation{ Univ. Lyon, ENS de Lyon, Univ. Claude Bernard, CNRS, Laboratoire de Physique, F-69342, Lyon.}

\maketitle

\tableofcontents

\renewcommand{\theequation}{S\arabic{equation}}
\renewcommand{\thefigure}{S\arabic{figure}}

\section{Stationary states and stability of (self-)driven crystals}
\subsection{Stationary states}
We consider the overdamped dynamics of an assembly of $N$ identical point particles at positions $\RR^\nu(t)$,
\begin{align}
   \nonumber
    \zeta \dot\RR^\nu &= 
    \sum_{\mu\neq \nu}\FF(\RR^\mu-\RR^\nu) \\
    &= 
    \sum_{j\neq i}\left[\FF\ind{A}(\RR^\mu-\RR^\nu) + \FF\ind{S}(\RR^\mu-\RR^\nu)\right]. \label{smeq:_overdamped}
\end{align}
Without loss of generality, we henceforth set  $\zeta = 1$.
We focus on homogeneous systems with pairwise interactions, and therefore assume that the  force $\FF$ only depends on the vector joining the two interacting particles. We decompose $\FF$ into
an antisymmetric part  $\FF\ind{A}$ satisfying $\FF\ind{A}(-\RR) = -\FF\ind{A}(\RR)$
and a symmetric part $\FF\ind{S}$ satisfying $\FF\ind{S}(-\RR) = \FF\ind{S}(\RR)$. We stress, as in the main text, that the effective force $\FF$ does not necessarily derive from an interaction potential, or free energy, and therefore escape the constraints imposed by Newton's third law. The system is driven far  from equilibrium and can continuously exchange both linear and angular momentum with its environment.

To address the stationarity of driven crystals, we consider that the particles initially form a  periodic lattice: $\RR^\nu(t=0) = \RR^{\nu}_0= n\ee_1 + m\ee_2$ where $\ee_1$ and $\ee_2$ are two independent vectors and  $\nu= (n, m)\in \mathbb{Z}^2$. Translational invariance then requires that the   forces acting on each particle are equal:
\begin{equation}
    \sum_{\mu\neq\nu} \FF(\RR^\mu - \RR^\nu) = \sum_{\mu'\neq 0} \FF(\RR^{\mu'} - \RR^{0}).
\end{equation} 
This relation implies that all  particles have the same velocity: $\dot\RR^\nu(t) = \dot\RR^0(t)$, and that their distances remain unchanged: $\RR^{\mu'}(t) - \RR^0(t) = \RR^{\mu'}_0 - \RR_0^0 =  \RR^{\mu'}_0 $ (we set $\RR_0^0 = \vec 0$). Furthermore, since $\RR^\mu_0$ and $-\RR^\mu_0$ both belong to the lattice, the contribution of antisymmetric forces vanish: $\sum_{\mu\neq 0} \FF\ind{A}(-\RR^\mu_0) = \vec 0$.
The dynamics of a perfect crystal then reduces to that of its center of mass and 
\begin{equation}
    \label{smeq:stationarity}
    \dot\RR^\nu = \dot\RR^0 = 
    \sum_{\mu\neq 0} \FF\ind{S}(\RR^\mu_0) \equiv \VV_0.
\end{equation}

\subsection{Angular modes of the interactions}
Crystalline structures are all stationary regardless of their specific spatial symmetry. 
We also note that when the interactions forces are antisymmetric, they cannot power the crystal translation. By contrast non-reciprocal symmetric forces may induce a steady rigid-body translation of the lattice. This  result holds in any dimension and is not merely restricted to the 2D case discussed in the main text.

Let us consider a 2D system following the dynamics of Eq.~\eqref{smeq:_overdamped}.
The polar coordinates associated with a vector $\RR$ are $(r, \theta)$ and we may represent this vector
by the complex number $\RR = r e^{i\theta}$. Provided that the angular symmetry of the interaction forces does not depends on distance, we can expand the force $\FF(\RR)$ as a sum over its angular modes,
\begin{equation} \label{smeq:angular_decompo}
    \FF(\RR = re^{i\theta}) = \sum_{n\in\mathbb{Z}} f_n(r) e^{i(n\theta - \alpha_n)}.
\end{equation}
The functions $f_n(r)$ are  real valued, and $\alpha_n$ is the phase of mode $n$.
The parity-odd modes (summation over $n$ odd) solely contribute to $\FF_A$ whereas the parity-even modes  ($n$ even) solely contribute to $\FF_S$.

To gain some intuition about the above angular decomposition, we briefly discuss some of the lowest order modes involved in prototypical examples of non-equilibrium soft condensed matter.
\begin{itemize}
\item {\bf Confined emulsions}. The case where the sole $n=2$ mode rules non-reciprocal interactions is realized by  emulsions driven in shallow channels. The droplet motion induces hydrodynamic perturbations quantitatively captured by dipolar flows~\cite{Desreumaux_2012,Beatus2006}. See Fig.~1b of the main text.
\item {\bf Sedimentation.} The hydrodynamic interactions governing the sedimentation of 2D lattices in 3D viscous fluids are accurately described by the so-called Stokeslet flow singularities, see e.g. Refs~\cite{Crowley1976,Chajwa2020}. Using the well-known Oseen tensor~\cite{Doi1988}, an external gravity force $-g\vec{\hat e}_y$ applied on a pointlike particle at the origin in a fluid of viscosity $\eta$ induces the following velocity field $\vec{V}(r, \theta)$ that in turn governs the hydrodynamic interactions $\FF(r, \theta)\propto \vec{V}(r, \theta)$,
\begin{equation}
    \vec{V}(r, \theta) = \frac{-g}{8\pi\eta r}\left(\vec{\hat e}_y + \frac{(\rr \cdot \vec{\hat e}_y) \rr}{r^2}\right)
    = \frac{-g}{16\pi\eta r}\left(3 e^{i\pi/2} +  e^{i(2\theta+\pi/2)}\right).
\end{equation}
This corresponds to the linear superposition of  
$n=0$ and $n=2$ non-reciprocal forces. $n=0$ forces correspond to interactions driving the particles along a constant direction while the $n=2$ mode has a dipolar symmetry. See Fig.~1a of the main text.
\item {\bf Active spinners}. As discussed e.g. in Refs.~\cite{Bililign_2021,Tan2021},  colloidal and biological spinners are primarily coupled by azimuthal flows corresponding to transverse drag forces where $n=1$ and $\alpha_1 = \pi/2$. See Fig.~1c of the main text.
\item {\bf Microswimmers.} 
Squirming active particles such as Janus colloids or active droplets all include $n=-1, n=1$ and $n=3$ modes in their field flow which chiefly determine hydrodynamic interactions~\cite{Lauga2020}.
The stresslet mode of the squirmer flow $\vec{V}$ (Ref.~\cite{Lauga2020}, Eqs.~(4.53)-(4.54)) is
\begin{align}
    \vec{V}(r, \theta) &= \frac{B_2}{2}\left\{
    \left(\frac{a^4}{r^4} - \frac{a^2}{r^2}\right)(3\cos^2\theta - 1) \vec{\hat{e}}_r +  \frac{a^4}{r^4}\sin(2\theta)   \vec{\hat{e}}_\theta
    \right\}\\
    &= \frac{B_2}{8} \left\{
    \left(\frac{2a^4}{r^4} - \frac{2a^2}{r^2}\right) e^{i\theta}
    + \left(\frac{a^4}{r^4} - \frac{3a^2}{r^2}\right) e^{-i\theta}
    + \left(\frac{5a^4}{r^4} - \frac{3a^2}{r^2}\right) e^{3i\theta}
    \right\},
\end{align}
where $a$ is the squirmer radius and $B_2$ the intensity of the stresslet. See Fig.~1d of the main text.
\item {\bf Higher order modes.} The case where $n=3$ could correspond to the quadrupolar coupling between electric dipoles having a fixed orientation in the plane. $n=4$  corresponds to forces having an hexapolar symmetry, etc. In the most general case bodies driven out of equilibrium by internal, and, or external drive interact via a number  of modes determined by the symmetry of the body shape and effective interactions. 
\end{itemize}

We now make a more technical comment and consider a given mode $\FF_n(r, \theta) = f_n(r) e^{i(n\theta-\alpha_n)}$. It is insightful to compute its divergence
\begin{align}
    \nabla\cdot \FF_n(r, \theta) = \frac{\cos[(n-1)\theta-\alpha_n]}{r} \left(n f_n(r) + rf_n'(r) \right).
\end{align}
We find that there exists two kinds of divergenceless forces ($\nabla\cdot \FF_n = 0$):
\begin{itemize}
    \item Transverse forces, realized when $n=1, \alpha_1 = \pm\pi/2$.
    \item Forces having a radial dependence $f_n(r) \propto r^{-n}$. These forces satisfy $\FF_n(\rr) = \nabla \phi_n(\rr)$ with $\phi_n \propto \cos((n-1)\theta-\alpha_n)/r^{n-1}$ (for $n\neq 1$). In this case $\phi_n$ is a solution of the Laplace equation $\Delta\phi_n(\rr) = 0$  for $\rr\neq 0$.
    $\phi_n$ and $\FF_n$ correspond to the $n$-th order of a 2D multipolar expansion. One can equivalently introduce the stream function $\psi_n(r, \theta) \propto \sin((n-1)\theta-\alpha_n)/r^{n-1}$ and check that $\FF_n(\rr) = \nabla^\perp \psi_n(\rr)$ (with $\nabla^\perp = (-\partial_y, \partial_x)$).
\end{itemize}

\subsection{Linear stability analysis}
To address the stability of  crystalline structures, we consider a small perturbation $\uu^\nu(t)$: $\RR^\nu(t) = \RR^\nu_0 + \VV_0 t + \uu^\nu(t)$. At the linear level, we consider without loss of generality this perturbation as a plane wave $\uu^\nu(t) = \uu(\qq, t) e^{-i\qq\cdot\RR^\nu_0}$ and compute its time evolution using  Eq.~\eqref{smeq:_overdamped},
\begin{align} \label{smeq:stability0}
    \dot\uu(\qq, t) &= 
    \sum_{\mu\neq 0} \left(e^{-i\qq\cdot\RR^\mu_0} - 1\right) 
    \nabla\left[\FF\ind{A} + \FF\ind{S}\right](\RR^\mu_0)\cdot\uu(\qq, t).
\end{align}
$\nabla\FF$ is the Jacobian matrix associated with $\FF$.
Using the fact that both $\RR^\mu_0$ and $-\RR^\mu_0$ belong to the lattice, and remembering the symmetries $\nabla\FF\ind{A}(-\RR^\mu_0) = \nabla\FF\ind{A}(\RR^\mu_0)$ and $\nabla\FF\ind{S}(-\RR^\mu_0) = -\nabla\FF\ind{S}(\RR^\mu_0)$ we obtain
\begin{align} \label{smeq:stability}
    \dot\uu(\qq, t) &= (M\ind{A} + i M\ind{S}) \cdot\uu(\qq, t)
\end{align}
with the real matrices
\begin{align}
    M\ind{A} &= -\sum_{\mu\neq 0}\left[1-\cos(\qq\cdot\RR^\mu_0)\right]\nabla\FF\ind{A}(\RR^\mu_0), \\
    M\ind{S} &= -\sum_{\mu\neq 0}  \sin(\qq\cdot\RR^\mu_0)\nabla\FF\ind{S}(\RR^\mu_0).
\end{align}

Let us detail the implications of Eq.~\eqref{smeq:stability} for a 2D system.
The eigenvalues of $M\ind{A}$ and $M\ind{S}$ are respectively denoted $\lambda\ind{A}^\pm$ and $\lambda\ind{S}^\pm$ and read
\begin{align}\label{smeq:stability_ev}
    \lambda\ind{A}^\pm &= \frac{1}{2} \left\{ \mathrm{tr}\, M_{A} \pm \left[(\mathrm{tr}\, M_{A})^2 - 4\det M_{A} \right]^{1/2} \right\}, &
    \lambda\ind{S}^\pm &=\frac{1}{2} \left\{ \mathrm{tr}\, M_{S} \pm \left[(\mathrm{tr}\, M_{S})^2 - 4\det M_{S} \right]^{1/2} \right\},
\end{align}
where an arbitrary branch cut is chosen to define the complex square root function.

The  stability of the system depends {\em a priori} on the details of the interactions.
However, we can show  that interactions violating Newton's third law are typically ineffective to stabilize periodic lattice against linear fluctuations: they either lead to an instability, or to wave propagation without damping (marginal stability). We can prove this result in the following two generic cases.
\begin{enumerate}
    \item {\bf Parity symmetric forces}. If $\Delta_S = (\mathrm{tr}\, M_{S})^2 - 4\det M_{S} \geq 0$, the eigenvalues $\lambda_S^\pm$ are real: following Eq.~\eqref{smeq:stability}, they lead to wave propagation without damping. On the other hand, if $\Delta_S < 0$, $\lambda_+$ and $\lambda_-$ have opposite imaginary parts: using Eq.~\eqref{smeq:stability} we see that one of these eigenvalues leads to an instability of the crystal structure.
    \item {\bf Divergenceless antisymmetric forces}. As show in the previous subsection, this includes generic tangential forces ($n=1, \alpha_1=\pi/2$) and the multipolar terms ($\FF(r, \theta) = e^{in\theta}/r^n$ for $n=3, 5, \dots$). In all these cases, $\lambda_A^\pm = \pm\sqrt{-\det M_A}$, and $\det M_A$ is either negative leading to a linear instability, or $\det M_A > 0$ and the system is marginally stable: plane waves  propagate without being damped.
\end{enumerate}

The discussion of whether we have instability or wave propagation depends on the microscopic details and cannot be done on the sole basis of symmetry considerations. The stability of a crystal with respect to dipolar forces ($n=2$, $\FF(\RR) = e^{2i\theta}/r^2$) was thoroughly investigated in Ref.~\cite{Desreumaux_2012}. It was found that the system is marginally stable that is to say that waves propagate without damping.
In the case of tangential forces ($n=1, \alpha_1=\pi/2$), we show in the next section that the continuous limit exhibits odd elasticity (coefficients given in Eqs.~\eqref{smeq:elast_coeffs}--\eqref{smeq:elast_coeffs2}): the stability analysis then reduces to Eq.~(5) in Ref.~\cite{Scheibner_2020}, and both instability and wave propagation are possible.

\section{Mechanics of non-reciprocal crystals: constitutive relations}

Our goal is now to relate the microscopic symmetries of the effective interaction forces into macroscopic constitutive relations for crystal escaping the constraints of Newton's third law.

\subsection{Deformation and stress tensors}
We first need  to define the displacement field $\uu(\rr) = (u_x(\rr), u_y(\rr))$, and the corresponding $2\times 2$ deformation tensor 
$D_{ij} = \frac{\partial u_i}{\partial x_j}$. $D$ has four components, which we express in the following basis (see Fig.~2b in the main text):
\begin{align}
    \tau^\delta &= \begin{pmatrix}1 & 0 \\ 0 & 1\end{pmatrix}, &
    \tau^\omega &= \begin{pmatrix}0 & -1 \\ 1 & 0\end{pmatrix}, &
    \tau^{s_1} &= \begin{pmatrix}1 & 0 \\ 0 & -1\end{pmatrix}, &
    \tau^{s_2} &= \begin{pmatrix}0 & 1 \\ 1 & 0\end{pmatrix}.
\end{align}
$\tau^\delta$ corresponds to a pure dilation, $\tau^\omega$ corresponds to a pure rotation, and $\tau^{s_1}$ and $\tau^{s_2}$ are two independent pure shear modes. In the following, we write $D = (D_\delta, D_\omega, D_{s_1}, D_{s_2})$ as a shortcut notation for
\begin{equation} \label{smeq:deform_compos}
    D_{ij} = D_\delta \tau^\delta_{ij} + D_\omega \tau^\omega_{ij}
    + D_{s_1} \tau^{s_1}_{ij} +  D_{s_2} \tau^{s_2}_{ij}.
\end{equation}

Internal forces in the system are characterized by the stress tensor $\sigma$. Considering two parts of our material, say $A$ and $B$ separated by an interface with normal vector $\nn^{A\to B}$, the force by unit length  $\TT^{B\to A}$ induced by  $B$ on $A$ is given by $T^{B\to A}_i = \sigma_{ij} n^{A\to B}_j$ (with summation on repeated indices). As we did for the deformation tensor, we use the decomposition $\sigma =(\sigma_\delta, \sigma_\omega, \sigma_{s_1}, \sigma_{s_2})$ as a shortcut for
\begin{equation} \label{smeq:stress_compos}
    \sigma_{ij} = \sigma_\delta \tau^\delta_{ij} + \sigma_\omega \tau^\omega_{ij}
    + \sigma_{s_1} \tau^{s_1}_{ij} +  \sigma_{s_2} \tau^{s_2}_{ij}.
\end{equation}

\subsection{Antisymmetric versus symmetric forces: stress-strain versus force-strain relations}
Antisymmetric forces ($\FF\ind{A}(-\RR) = -\FF\ind{A}(\RR)$) correspond to angular modes with $n$ odd (Eq.~\eqref{smeq:angular_decompo}). We previously showed that in a periodic lattice, the sum of antisymmetric  forces on a given particle cancel out. As in equilibrium systems, this results implies that the lowest order constitutive relation relates stresses and deformations. At linear order
\begin{equation}
    \sigma_{ij} = \Pi_{ij} + K_{ijkl} D_{kl}.
    \label{Eq:stressstrain}
\end{equation}
where $\Pi$ is a possible pre-stress in the absence of deformations.
The goal of subsection~\ref{ss:asym} is to explicitly construct this  relation for hexagonal lattices.

Symmetric forces ($\FF\ind{S}(-\RR) = \FF\ind{S}(\RR)$) correspond to angular modes with $n$ even (Eq.~\eqref{smeq:angular_decompo}). In this case, a deformation of the crystal can induce a net body force  $\mathcal{F}$, which at linear order results in the atypical constitutive relation
\begin{equation}
    \mathcal{F}_{i} = \mathcal{F}_{i}^0 + A_{ikl} D_{kl}.
\end{equation}
Finding the force-deformation relation induced by symmetric modes is the aim of subsection~\ref{ss:sym}.

\begin{figure}
    \centering
    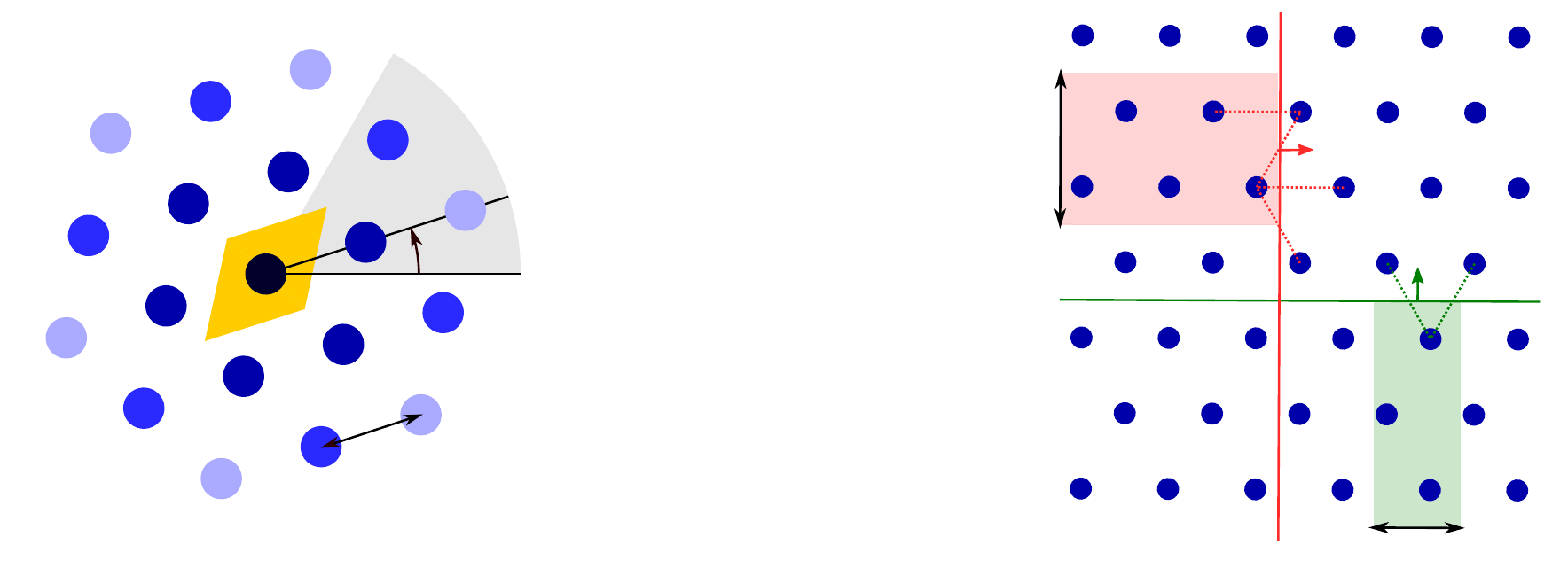
    \caption{
    (a) Basic setup to compute the stress tensor generated by parity-odd forces from Eq.~\eqref{eq:micro_cauchy}; or the net force generated by parity-even forces from Eq.~\eqref{smeq:f_even}. The test particle is represented in black, its nearest neighbors in dark blue, and particles further away in lighter shades of blue. The lattice spacing is $a$ and the lattice orientation is $\Theta$. The unit cell of area $V=\sqrt{3}a^2/2$ is represented in orange, and the angular section $\mathbb{H}_{/6}$ corresponding to angle $\theta\in[0,\pi/3)$ is represented in gray.
    (b) The stress tensor can be written $\sigma = (\TT^x, \TT^y)$ with $\TT^x$ (resp. $\TT^y$) the force applied by the particles outside of the black rectangle to the particles inside across the right segment (resp. across the top segment) of the box.
    (c) Setup for the alternative derivation of the stress tensor generated by parity-odd forces (subsection~\ref{ss:alternative}). 
    A cut normal to $\hat{\vec{x}}$ is represented in red. By definition $\TT^x$ is the force by unit length exerted by the particles on the right of this cut onto the particles on the left. We compute the force applied on a stripe of length $\ell_x$ (red stripe). The interactions between nearest neighbors are shown with dotted red lines. A similar discussion holds for the cut normal to $\hat{\vec{y}}$ represented in green.
    }
    \label{fig:forces_stress}
\end{figure}

\subsection{Parity-odd interactions: odd elasticity and beyond} \label{ss:asym}
\subsubsection{Computation of the stress tensor}
In this subsection, we consider a 2D hexagonal lattice, denote  $a$ the lattice spacing and consider particles interacting via  pairwise antisymmetric forces $\FF\ind{A}(-\RR) = -\FF\ind{A}(\RR)$ only. The total force on any given particle of the crystal vanishes by symmetry (even in the presence of  linear deformations). 
Our goal is now to describe the internal stresses. At linear order in the deformations (see Eq.~\eqref{smeq:deform_compos} for the decomposition), we expect an affine relation
\begin{equation}
    \begin{pmatrix}
    \sigma_\delta \\ \sigma_\omega \\ \sigma_{s_1} \\ \sigma_{s_2}
    \end{pmatrix}
    = \begin{pmatrix}
    P_\delta \\ T_\omega \\ S_1 \\ S_2
    \end{pmatrix}
    + \begin{pmatrix}\ast & \ast & \ast & \ast \\ \ast & \ast & \ast & \ast \\ \ast & \ast & \ast & \ast \\ \ast & \ast & \ast & \ast\end{pmatrix}
    \begin{pmatrix}
    D_\delta \\ D_\omega \\ D_{s_1} \\ D_{s_2}
    \end{pmatrix}.
\end{equation}
The vector on the r.h.s. corresponds to the analog of pre-stresses in equilibrium solids, and the $4\times 4$ matrix accounts for the linear relation between the stress and strain ($K$ tensor in Eq.~\eqref{Eq:stressstrain}).

Following, e.g. Refs~\cite{Irving_1950,Yang2012,Klymko2017}, we use the Irving-Kirkwood virial formula to write the Cauchy stress tensor. For a lattice with one atom per unit cell, the stress is given by
\begin{equation} \label{eq:micro_cauchy}
    \sigma_{ij} = \frac{1}{2V} \sum_\nu F_i(\RR^\nu) R_j^\nu,
\end{equation}
where we ignore the so-called kinetic contributions having athermal systems in mind. 
$V$ is the volume of the unit cell  ($V=\sqrt{3}a^2/2$ for a hexagonal lattice with spacing $a$), and the summation is performed over all particles  (excluding the one at the origin). For simplicity we dropped the 'A' subscript of $\FF$. The sum of Eq.~\eqref{eq:micro_cauchy} is illustrated in Fig.~\ref{fig:forces_stress}a.

In the following, we focus on forces that are expressed in polar coordinates as
\begin{equation}
    \FF(R, \theta) = f(R) \Gg(\theta).
\end{equation}
In this case, the stress tensor in the absence of deformation reads
\begin{equation} \label{smeq:stress_nodef0}
    \sigma_{ij}^0 = \frac{1}{2V} \sum_\nu f(R^\nu) g_i(\theta^\nu) R_j^\nu.
\end{equation}
In response to a deformation mapping $\RR^\nu$ to $\RR^\nu+\delta\RR^\nu$, the volume of the unit cell is changed by $\delta V$, and an additional stress $\delta \sigma$ builds up. Applying the chain rule we find:
\begin{equation}\label{smeq:stress_def0}
	\delta \sigma_{ij} = \frac{1}{2V} \sum_\nu \left[f'(R^\nu) g_i(\theta^\nu)R_j^\nu \delta R^\nu + f(R^\nu) g_i'(\theta^\nu) R_j^\nu \delta\theta^\nu + f(R^\nu) g_i(\theta^\nu) \delta R_j^\nu -  f(R^\nu) g_i(\theta^\nu) R_j^\nu \frac{\delta V}{V}  \right].
\end{equation}
This seemingly complex formula actually provides a simple insight in the generation of stresses in non-reciprocal lattices.

\subsubsection{Using the symmetry of the hexagonal lattice}
\textbf{No deformation.}  In the case of the hexagonal lattice, we can exploit the rotational symmetry of order $6$. As illustrated in Fig.~\ref{fig:forces_stress}a,
the summation in Eq.~\eqref{smeq:stress_nodef0} can then be limited to the set of particles inside the angular sector $\theta^\nu\in [0, \pi/3)$ which we note $\mathbb{H}_{/6}$,   provided that for each particle in $\mathbb{H}_{/6}$ we also account for the effect of the $5$ other particles located at angles $\theta^{\nu,k} = \theta^{\nu}+k \frac{\pi}{3}$ for $k=1, \dots, 5$. Eq.~\eqref{smeq:stress_nodef0} is then recast as
\begin{equation}\label{smeq:stress_nodef1}
\sigma_{ij}^0 = \frac{1}{2V} \sum_{\nu\in\mathbb{H}_{/6}} R^\nu f(R^\nu) \sum_{k=0}^5 g_i(\theta^{\nu,k}) \left[e^{i\theta^{\nu,k}}\right]_j
\end{equation}
where $[z]_1 = \mathrm{Re}(z)$ is the real part of the complex number $z$ and $[z]_2 = \mathrm{Im}(z)$ is its imaginary part. 
In the next subsection, we will show that the sum over $k$ can be evaluated explicitly for a given choice of $\Gg(\theta)$. 
This will give us the four components (dilation, odd stress and shear components) of the stress tensor from both the symmetries of the hexagonal lattice and the symmetries of the interaction.

Hexagonal lattices with nearest-neighbor interactions (`nn') are of particular interest. In this case,  the sum over $\nu$ reduces to a single term ($R^\nu = a$). Eq.~\eqref{smeq:stress_nodef1} then becomes
\begin{equation}\label{smeq:stress_nodef1_nn}
    \sigma_{ij}^{0} \overset{nn}{=} \frac{f(a)}{\sqrt{3} a} \sum_{k=0}^5 g_i\left(\Theta + k\frac{\pi}{3}\right) \left[e^{i\left(\Theta + k\frac{\pi}{3}\right)}\right]_j
\end{equation}
where we used $V=\sqrt{3}a^2/2$ and called $\Theta$ the orientation of the lattice so that the first shell of neighbors corresponds to angles $\Theta + k\frac{\pi}{3}$ (see Fig.~\ref{fig:forces_stress}).
Expressions similar to Eq.~\eqref{smeq:stress_nodef1_nn} can be easily obtained from Eqs.~\eqref{smeq:stress_nodef1}-\eqref{smeq:stress_shear1} below, for the stress tensor in a deformed system with nearest-neighbors interactions only.

\textbf{Dilation.} A dilation of amplitude $D_\delta$ corresponds to the complex map $z\mapsto (1+D_\delta) z$. This implies $\delta\RR = D_\delta\RR$, $\delta R = D_\delta R$ and $\delta\theta = 0$. 
In addition, the dilation of the unit cell at linear order is $\delta V / V = 2 D_\delta$. We stress that this is the only mode of deformation for which $\delta V \neq 0$. 
It is then straightforward to derive the variations of stress tensor from Eq.\eqref{smeq:stress_def0}:
\begin{align}\label{smeq:stress_dil1}
&= \frac{D_\delta}{2V} \sum_{\nu\in\mathbb{H}_{/6}} R^\nu \left[
R^\nu f'(R^\nu) - f(R^\nu)  \right] \sum_{k=0}^5 g_i(\theta^{\nu,k}) \left[e^{i\theta^{\nu,k}}\right]_j.
\end{align}

\textbf{Rotation.} The simplest way to evaluate the stress  resulting from a rotation of angle $D_\omega$ consists in expanding Eq.~\eqref{smeq:stress_nodef1} at linear order in $D_\omega$ when $\theta^{\nu,k}\mapsto \theta^{\nu,k}+D_\omega$. This leads to
\begin{align}\label{smeq:stress_rot1}
\delta \sigma_{ij}^\mathrm{rot} = \frac{D_\omega}{2V} \sum_{\nu\in\mathbb{H}_{/6}} R^\nu f(R^\nu) 
 \sum_{k=0}^5  \left\{ g_i' (\theta^{\nu,k}) \left[e^{i\theta^{\nu,k}}\right]_j
+  g_i(\theta^{\nu,k}) \left[i e^{i\theta^{\nu,k}}\right]_j
  \right\}.
\end{align}

\textbf{Pure shear.}  A pure shear of angle $\gamma$ and amplitude $D_s$ corresponds to the map $z\mapsto z + D_s e^{2i\gamma} z^\ast$.
This implies that $\delta\rr = D_s e^{2i\gamma} r e^{-i\theta}$, $\delta r = r \cos[2(\gamma-\theta)]$ and $\delta\theta = \sin[2(\gamma-\theta)]$.
Using Eq.\eqref{smeq:stress_def0}, we find the stress variations
\begin{multline}\label{smeq:stress_shear1}
\delta \sigma_{ij}^\mathrm{shear} = \frac{D_s}{2V} \sum_{\nu\in\mathbb{H}_{/6}} \bigg\{
(R^\nu)^2 f'(R^\nu)  \sum_{k=0}^5   g_i(\theta^{\nu,k}) \cos[2(\gamma-\theta^{\nu,k})] \left[e^{i\theta^{\nu,k}}\right]_j \\
+ R^\nu f(R^\nu)  \sum_{k=0}^5  \left(
	g_i'(\theta^{\nu,k}) \sin[2(\gamma-\theta^{\nu,k})] \left[e^{i\theta^{\nu,k}}\right]_j
	+ g_i(\theta^{\nu,k}) \left[e^{i\left(2\gamma - \theta^{\nu,k}\right)}\right]_j
\right)
  \bigg\}.
\end{multline}

\subsubsection{Angular modes of interactions}
Having derived the general expressions Eqs.~\eqref{smeq:stress_nodef1}-\eqref{smeq:stress_shear1}, we focus on a given angular mode (see Eq.~\eqref{smeq:angular_decompo}), $\Gg(\theta) = e^{i(n\theta-\alpha_n)}$ with $n$ odd. All summations over $k$ in Eqs.~\eqref{smeq:stress_nodef1}-\eqref{smeq:stress_shear1} involve the basic sum
\begin{equation}
    d_m(\theta)\equiv \frac{1}{6}\sum_{k=0}^5 e^{im\left(\theta+k\frac{\pi}{3}\right)} = e^{im\theta} \sum_{p=-\infty}^\infty \delta_{m, 6p} 
    \label{eq:trick}
\end{equation}
which is nonzero only if $m$ is a multiple of $6$.

In practice, it is convenient to compute the forces (by unit length)
\begin{equation}
  \TT^x = \sigma_{xx} + i\sigma_{yx}  
\end{equation}
 and 
 \begin{equation}
     \TT^y = \sigma_{xy} + i \sigma_{yy}
 \end{equation} 
 acting on surfaces with normal vectors $\hat{\vec{x}}$ and $\hat{\vec{y}}$. This representation allows a more compact presentation of our results.

This representation makes it clear that two symmetries are of particular interest:
\begin{enumerate}
    \item If $\TT^y = i\TT^x$, the stress tensor is made of a dilation component of amplitude $\sigma_\delta = \mathrm{Re}(\TT^x)$
and an odd stress component of amplitude $\sigma_\omega = \mathrm{Im}(\TT^x)$. 

\item If $\TT^y = -i\TT^x$, the stress tensor is a pure shear matrix made of a shear 1 component of amplitude $\sigma_{s1} = \mathrm{Re}(\TT^x)$ and a shear 2 component of amplitude $\sigma_{s1} =\mathrm{Im}(\TT^x)$.
\end{enumerate}

We are now ready to reformulate Eqs.~\eqref{smeq:stress_nodef1}-\eqref{smeq:stress_shear1} in the basic case where $\Gg(\theta) = e^{i(n\theta-\alpha_n)}$:
\begin{align}
\label{smeq:stress_nodef2}
    \begin{pmatrix}
    \TT^x \\ \TT^y
    \end{pmatrix}^0
    &= \frac{3}{2V}e^{-i\alpha_n}  \sum_{\nu\in\mathbb{H}_{/6}} R^\nu f(R^\nu)
    \left\{
    d_{n-1}(\theta^\nu)\begin{pmatrix} 1 \\ i \end{pmatrix}
    +d_{n+1}(\theta^\nu)\begin{pmatrix} 1 \\ -i \end{pmatrix}
    \right\}, \\
    \label{smeq:stress_dilat2}
    \begin{pmatrix}
    \delta\TT^x \\ \delta\TT^y
    \end{pmatrix}^\mathrm{dilat}
    &= \frac{3D_\delta}{2V}e^{-i\alpha_n}  \sum_{\nu\in\mathbb{H}_{/6}} R^\nu \left[
    R^\nu f'(R^\nu) - f(R^\nu)  \right]
    \left\{
    d_{n-1}(\theta^\nu)\begin{pmatrix} 1 \\ i \end{pmatrix}
    + d_{n+1}(\theta^\nu) \begin{pmatrix} 1 \\ -i \end{pmatrix}
    \right\}, \\
    \begin{pmatrix}
    \label{smeq:stress_rot2}
    \delta\TT^x \\ \delta\TT^y
    \end{pmatrix}^\mathrm{rot}
    &= \frac{3D_\omega}{2V} ie^{-i\alpha_n}  \sum_{\nu\in\mathbb{H}_{/6}} R^\nu f(R^\nu)
    \left\{
    (n-1)d_{n-1}(\theta^\nu)\begin{pmatrix} 1 \\ i \end{pmatrix}
    + (n+1)d_{n+1}(\theta^\nu)\begin{pmatrix} 1 \\ -i \end{pmatrix}
    \right\}, \\
    \begin{pmatrix}
    \delta\TT^x \\ \delta\TT^y
    \end{pmatrix}^\mathrm{shear}
    &= \frac{3D_s}{4V}e^{-i\alpha_n}  \sum_{\nu\in\mathbb{H}_{/6}} R^\nu \bigg\{
    \left[R^\nu f'(R^\nu) + (n+2) f(R^\nu)  \right] d_{n-1}(\theta^\nu) e^{2i\gamma}\begin{pmatrix} 1 \\ -i \end{pmatrix}\nonumber \\
    &+  \left[R^\nu f'(R^\nu) + n f(R^\nu)  \right]  d_{n-3}(\theta^\nu)e^{2i\gamma}\begin{pmatrix} 1 \\ i \end{pmatrix}
    + \left[R^\nu f'(R^\nu) - n f(R^\nu)  \right] d_{n+3}(\theta^\nu) e^{-2i\gamma} \begin{pmatrix} 1 \\ -i \end{pmatrix} \label{smeq:stress_shear2}  \\
    &+ \left[R^\nu f'(R^\nu) - (n-2) f(R^\nu)  \right]   d_{n+1}(\theta^\nu) e^{-2i\gamma}\begin{pmatrix} 1 \\ i \end{pmatrix}
    \bigg\}, \nonumber
\end{align}
where the $d_m$ coefficients are given by Eq.~\eqref{eq:trick}.
It is worth noting that $\begin{pmatrix} 1 \\ i \end{pmatrix}$ vectors correspond to pressure and odd stress terms (depending on the phase), while $\begin{pmatrix} 1 \\ -i \end{pmatrix}$ vectors correspond to  shear stresses. 

Before investigating the three generic cases $n=6p+1$, $n=6p+3$ and $n=6p+5$ (corresponding to the three possible parities of $n$ modulo $6$), we first gain  some intuition by studying in details the important case $n=1$. \\

\textbf{Isotropic forces $n=1$.}
Both longitudinal and transverse isotropic forces correspond to first order term ($n=1$) in the expansion Eq.~\eqref{smeq:angular_decompo}. They take the form 
\begin{equation} \label{smeq:_isotropic}
    \FF(re^{i\theta}) = f(r) e^{i(\theta - \alpha_1)}, 
\end{equation}
where $f(r)$ is the radial dependence, and $\alpha_1$ the phase. The case $\alpha_1=0$ corresponds to longitudinal forces that satisfy Newton's third law, and the case $\alpha_1 = \pm\pi/2$ corresponds to pure transverse forces.

When $n=1$, Eqs.~\eqref{smeq:stress_nodef2}-\eqref{smeq:stress_shear2} reduce to
\begin{align}
\label{smeq:stress_nodef_1}
    &\begin{pmatrix}
    \TT^x \\ \TT^y
    \end{pmatrix}^0
    = \begin{pmatrix} 1 \\ i \end{pmatrix}e^{-i\alpha_1}\, \frac{3}{2V}  \sum_{\nu\in\mathbb{H}_{/6}} R^\nu f(R^\nu), \\
    \label{smeq:stress_dilat_1}
    &\begin{pmatrix}
    \delta\TT^x \\ \delta\TT^y
    \end{pmatrix}^\mathrm{dilat}
    = D_\delta \begin{pmatrix} 1 \\ i \end{pmatrix}e^{-i\alpha_1}\, \frac{3}{2V}  \sum_{\nu\in\mathbb{H}_{/6}} R^\nu \left[
    R^\nu f'(R^\nu) - f(R^\nu)  \right], \\
        \label{smeq:stress_rot_1}
    &\begin{pmatrix}
    \delta\TT^x \\ \delta\TT^y
    \end{pmatrix}^\mathrm{rot}
    = 
    \begin{pmatrix}
    0 \\ 0
    \end{pmatrix}, \\
        \label{smeq:stress_shear_1}
    &\begin{pmatrix}
    \delta\TT^x \\ \delta\TT^y
    \end{pmatrix}^\mathrm{shear}
    = D_s \begin{pmatrix} 1 \\ -i \end{pmatrix}e^{i(2\gamma-\alpha_1)} \frac{3}{4V}  \sum_{\nu\in\mathbb{H}_{/6}} R^\nu 
    \left[R^\nu f'(R^\nu) + 3f(R^\nu)  \right].
\end{align}
In the absence of deformations, the pre-stress has a dilation component ($\propto \cos\alpha_1$) and a rotation component ($\propto \sin\alpha_1$). These two components are also found in response to a dilation deformation. Due to the isotropy of the interactions, the system is insensitive to rotations. Finally, a shear deformation translates into a shear stress with a component corresponding to the same shear mode ($\propto \cos\alpha_1$) and a component on the orthogonal shear mode ($\propto \sin\alpha_1$).

All in all, the constitutive relation of a crystal assembled from units interacting via  isotropic forces is then given by
\begin{equation} \label{smeq:stress_deform_iso}
    \begin{pmatrix}
    \sigma_\delta \\ \sigma_\omega \\ \sigma_{s_1} \\ \sigma_{s_2}
    \end{pmatrix}
    = \begin{pmatrix}
    \tilde P\cos\alpha_1 \\ -\tilde P\sin\alpha_1 \\ 0 \\ 0
    \end{pmatrix}
    + \begin{pmatrix} \tilde B\cos\alpha_1 & 0 & 0 & 0 \\ -\tilde B\sin\alpha_1 & 0 & 0 & 0 \\
    0 & 0 & \tilde\mu \cos\alpha_1 & -\tilde\mu \sin\alpha_1 \\ 0 & 0 & \tilde\mu \sin\alpha_1 & \tilde\mu \cos\alpha_1 \end{pmatrix}
    \begin{pmatrix}
    D_\delta \\ D_\omega \\ D_{s_1} \\ D_{s_2}
    \end{pmatrix}.
\end{equation}
We recall the meaning of all the elastic parameters. $B=\tilde B\cos\alpha_1$ and $\mu=\tilde\mu \cos\alpha_1$ are the bulk and shear moduli. They stem from longitudinal forces only. The coefficients $A=-\tilde B\sin\alpha_1$ and $K^o=-\tilde\mu \sin\alpha_1$ are the two odd elasticity coefficients, they couple respectively a dilation to a torque and a shear deformation to an orthogonal shear stress~\cite{Scheibner_2020}. They emerge from the sole effect of microscopic transverse forces.

The elastic parameters are given by
\begin{align} \label{smeq:elast_coeffs}
    \tilde P &= \frac{\sqrt{3}}{a^2} \sum_{\nu\in\mathbb{H}_{/6}} R^\nu f(R^\nu) \overset{nn}{=} \frac{\sqrt{3} f(a)}{a}, \\
    \tilde B &= \frac{\sqrt{3} f(a)}{a} \sum_{\nu\in\mathbb{H}_{/6}} R^\nu \left[
    R^\nu f'(R^\nu) - f(R^\nu)  \right]  \overset{nn}{=} \frac{\sqrt{3}}{a}\left[a f'(a) - f(a)\right], \\
    \tilde\mu &= \frac{\sqrt{3}}{2a^2}\sum_{\nu\in\mathbb{H}_{/6}} R^\nu 
    \left[R^\nu f'(R^\nu) + 3f(R^\nu)  \right]  \overset{nn}{=} \frac{\sqrt{3}}{2a}\left[a f'(a) + 3f(a)\right].
    \label{smeq:elast_coeffs2}
\end{align}
where $\overset{nn}{=}$ gives the expression for nearest-neighbor interactions. 
We are now equipped to address the three general cases.

\textbf{Non-reciprocal interactions of orders $n=6p+1$.} The angular orders $n=6p+1$ generalize the case $n=1$. Eqs.~\eqref{smeq:stress_nodef_1}, \eqref{smeq:stress_dilat_1} and \eqref{smeq:stress_shear_1} have similar expressions with an extra factor $e^{i(n-1)\theta^\nu}$ in the sum. The main difference is that for $n\neq 1$, the system is not isotropic:
a rotation induces a pressure and an odd stress given by
\begin{equation}
    \begin{pmatrix}
    \delta\TT^x \\ \delta\TT^y
    \end{pmatrix}^\mathrm{rot}
    = \frac{3D_\omega}{2V} (n-1) i\begin{pmatrix} 1 \\ i \end{pmatrix} e^{-i\alpha_n}  \sum_{\nu\in\mathbb{H}_{/6}} R^\nu f(R^\nu) e^{i(n-1)\theta^\nu}.
\end{equation}

The structure of the stress-deformation relation is 
\begin{equation} \label{smeq:stress_1}
    \begin{pmatrix}
    \sigma_\delta  \\ \sigma_\omega  \\ \sigma_{s_1} \\ \sigma_{s_2}
    \end{pmatrix}_{(n=6p+1)}
    = \begin{pmatrix}
    P  \\ T  \\ 0 \\ 0
    \end{pmatrix}
    + 
    \begin{pmatrix}
    B & \Lambda & 0 & 0 \\
    A & \Gamma & 0 & 0 \\
    0 & 0 & \mu & K^o \\
    0 & 0 & -K^o & \mu
    \end{pmatrix}
    \begin{pmatrix}
    D_\delta  \\ D_\omega  \\ D_{s_1} \\ D_{s_2}
    \end{pmatrix}.
\end{equation}
For simplicity, we give the expressions of the coefficients in the nearest-neighbors case only:
\begin{align}
    P &\overset{nn}{=} \frac{\sqrt{3} f(a)}{a}\cos[(n-1)\Theta - \alpha_n], &
    T &\overset{nn}{=} \frac{\sqrt{3} f(a)}{a}\sin[(n-1)\Theta - \alpha_n], \\
    B &\overset{nn}{=} \frac{\sqrt{3}}{a}\left[a f'(a) - f(a)\right]\cos[(n-1)\Theta - \alpha_n], &
    A &\overset{nn}{=} \frac{\sqrt{3}}{a}\left[a f'(a) - f(a)\right]\sin[(n-1)\Theta - \alpha_n], \\
    \Lambda &\overset{nn}{=} -(n-1)\frac{\sqrt{3} f(a)}{a}\sin[(n-1)\Theta - \alpha_n], &
    \Gamma &\overset{nn}{=} (n-1)\frac{\sqrt{3} f(a)}{a}\cos[(n-1)\Theta - \alpha_n], \\
    \mu &\overset{nn}{=}  \frac{\sqrt{3}}{2a}\left[a f'(a) + (n+2)f(a)\right]\cos[(n-1)\Theta - \alpha_n], &
    K^o &\overset{nn}{=}  \frac{\sqrt{3}}{2a}\left[a f'(a) + (n+2)f(a)\right]\sin[(n-1)\Theta - \alpha_n], \\
\end{align}
where $\Theta$ is the orientation of the lattice. \\

\textbf{Non-reciprocal interactions of orders $n=6p+3$.} Expressions of the stress for angular symmetries $n=6p+3$ can be obtained by collecting the terms $d_{n-3}$ and $d_{n+3}$ in Eqs.~\eqref{smeq:stress_nodef2}-\eqref{smeq:stress_shear2}. The most striking feature is that the system responds to shear deformations only. Moreover, in response to a shear deformation, all four stress modes are activated. The constitutive relation reads
\begin{equation} \label{smeq:stress_3}
    \begin{pmatrix}
    \sigma_\delta  \\ \sigma_\omega  \\ \sigma_{s_1} \\ \sigma_{s_2}
    \end{pmatrix}_{(n=6p+3)}
    = \begin{pmatrix}
    0 \\ 0  \\ 0 \\ 0
    \end{pmatrix}
    + 
    \begin{pmatrix}
    0 & 0 & C_1 & -C_2 \\
    0 & 0 & C_2 & C_1 \\
    0 & 0 & \mu_1 &  \mu_2 \\
    0 & 0 & \mu_2 & -\mu_1
    \end{pmatrix}
    \begin{pmatrix}
    D_\delta  \\ D_\omega  \\ D_{s_1} \\ D_{s_2}
    \end{pmatrix}.
\end{equation}
The lower-right $2\times 2$ block corresponds to a standard anisotropic shear elasticity with eigenvalues $\pm\tilde\mu$ (one direction is stable and the other one is unstable), it includes no odd elastic contribution.

The coefficients in the nearest-neighbors case are
\begin{align}
    C_1 &\overset{nn}{=} \frac{\sqrt{3}}{2a} \left[a f'(a) + nf(a)\right]\cos[(n-3)\Theta - \alpha_n], &
    C_2 &\overset{nn}{=} \frac{\sqrt{3}}{2a} \left[a f'(a) + nf(a)\right]\sin[(n-3)\Theta - \alpha_n], \\
    \mu_1 &\overset{nn}{=} \frac{\sqrt{3}}{2a} \left[a f'(a) - nf(a)\right]\cos[(n+3)\Theta - \alpha_n], &
    \mu_2 &\overset{nn}{=} \frac{\sqrt{3}}{2a} \left[a f'(a) - nf(a)\right]\sin[(n+3)\Theta - \alpha_n]. 
\end{align}
\\

\textbf{Non-reciprocal interactions of order $n=6p+5$.} The main peculiarity of solids with interactions having an angular symmetry of order $n=6p+5$ is that they support pure shear stresses even in the absence of any external deformations. Moreover, such solids produce isotropic stresses and torques when sheared, and shear stresses when dilated or rotated.

From Eqs.~\eqref{smeq:stress_nodef2}-\eqref{smeq:stress_shear2}, the characteristic relation is
\begin{equation} \label{smeq:stress_5}
    \begin{pmatrix}
    \sigma_\delta  \\ \sigma_\omega  \\ \sigma_{s_1} \\ \sigma_{s_2}
    \end{pmatrix}_{(n=6p+5)}
    = \begin{pmatrix}
    0 \\ 0  \\ S_1 \\ S_2
    \end{pmatrix}
    + 
    \begin{pmatrix}
    0 & 0 & E_1 & E_2 \\
    0 & 0 & E_2 & -E_1 \\
    G_1 & G_2' & 0 & 0 \\
    G_2 & G_1' & 0 & 0
    \end{pmatrix}
    \begin{pmatrix}
    D_\delta  \\ D_\omega  \\ D_{s_1} \\ D_{s_2}
    \end{pmatrix},
\end{equation}
and the coefficients in the nearest-neighbors case are
\begin{align}
    S_1 &\overset{nn}{=} \frac{\sqrt{3} f(a)}{a}\cos[(n+1)\Theta - \alpha_n], &
    S_2 &\overset{nn}{=} \frac{\sqrt{3} f(a)}{a}\sin[(n+1)\Theta - \alpha_n], \\
    G_1 &\overset{nn}{=} \frac{\sqrt{3}}{a}\left[a f'(a) - f(a)\right]\cos[(n+1)\Theta - \alpha_n], &
    G_2 &\overset{nn}{=} \frac{\sqrt{3}}{a}\left[a f'(a) - f(a)\right]\sin[(n+1)\Theta - \alpha_n], \\
    G_2' &\overset{nn}{=} -(n+1)\frac{\sqrt{3} f(a)}{a}\sin[(n+1)\Theta - \alpha_n], &
    G_1' &\overset{nn}{=} (n+1)\frac{\sqrt{3} f(a)}{a}\cos[(n+1)\Theta - \alpha_n], \\
    E_1 &\overset{nn}{=} \frac{\sqrt{3}}{2a} \left[a f'(a) - (n-2)f(a)\right] \cos[(n+1)\Theta - \alpha_n], &
    E_2 &\overset{nn}{=} \frac{\sqrt{3}}{2a} \left[a f'(a) - (n-2)f(a)\right] \sin[(n+1)\Theta - \alpha_n].
\end{align}

\subsubsection{Alternative derivation} \label{ss:alternative}
In this section, we provide an alternative derivation of the results of the previous subsections. 
Instead of relying on the Irving-Kirkwood formula, Eq.~\eqref{eq:micro_cauchy}, we  start from the bare mechanical definition of the Cauchy stress tensor.
We write $\sigma = (\TT^x, \TT^y)$ i.e. the matrix made of the two column vectors $\TT^x$ and $\TT^y$ where $\TT^x$ is the force per unit length exerted along a vertical cut by the particles on the right of the cut on those on the left, and $\TT^x$ is the force per unit length exerted along an horizontal cut by the particles at the top on those at the bottom,
see Fig.~\ref{fig:forces_stress}c.

We consider a hexagonal lattice with an horizontal principal axis ($\Theta=0$).
For simplicity, we restrict ourselves to a nearest neighbors analysis%
\footnote{Note that this  method can be extended to additional shells of next-nearest neighbors. For instance, the second shell of neighbors (at distance $\sqrt{3}a$) can be incorporated into $\TT^x$ as $3\FF\left(\sqrt{3}a e^{i\frac{\pi}{6}}\right) + 3\FF\left(\sqrt{3}a e^{i\frac{\pi}{6}}\right)$. However, the counting procedure and the computations  become increasingly tedious when considering extra particles. The Irving-Kirkwood method facilitates the calculation when the sum is performed beyond the n.n. approximation Eq.~\eqref{eq:micro_cauchy}.}
in which the two forces $\TT^x$ and $\TT^y$ read
\begin{align} \label{smeq:TxTy}
    \TT^x &= \frac{1}{\ell_x} \left[ 2\FF(a) + \FF\left(a e^{i\frac{\pi}{3}}\right) +  \FF\left(a e^{-i\frac{\pi}{3}}\right)\right], &
    \TT^y &= \frac{1}{\ell_y} \left[ \FF\left(a e^{2i\frac{\pi}{3}}\right) +  \FF\left(a e^{-2i\frac{\pi}{3}}\right)\right].
\end{align}
where the 2D positions are expressed as complex numbers and $a$ is the lattice spacing.
$\ell_x = \sqrt{3}a$ and $\ell_y = a$ are the units of length associated with the periodicity along the cuts normal to $x$ and $y$ (see Fig.~\ref{fig:forces_stress}c).

We now consider an interaction of the form
\begin{equation} \label{smeq:fr_other}
    \FF(r e^{i\theta}) = f(r) e^{i(n\theta-\alpha_n)}
\end{equation}
where $n$  is an odd integer, and evaluate Eq.~\eqref{smeq:TxTy} first in an undeformed, then in a deformed solid. \\

\noindent \textbf{No deformation.}  Combining Eqs.~\eqref{smeq:TxTy} and~\eqref{smeq:fr_other} and remembering that $n$ is odd, we find
\begin{align} \label{smeq:nodeform2}
    \begin{pmatrix}
     \TT^x \\\TT^y
    \end{pmatrix}^0
    = \frac{\sqrt{3}f(a)e^{-i\alpha_n}}{a}
    \left\{\delta_{n, 6p+1} \begin{pmatrix}1 \\ i\end{pmatrix} 
    + \delta_{n, 6p+5} \begin{pmatrix}1 \\ -i\end{pmatrix}\right\}
\end{align}
where $\delta_{n, 6p+1}=1$ if $n=1$  modulo $6$, and $\delta_{n, 6p+1}=0$ otherwise (and \textit{mutatis mutandis} for $\delta_{n, 6p+5}$).
This results holds for an hexagonal lattice of orientation $\Theta = 0$. We now address the case of an arbitrary orientation $\Theta$. Two ingredients need to be taken into account.
\begin{enumerate}
    \item Under a rotation of the system by an angle $\Theta$, the stress tensor transforms as a twice-contravariant tensor. Two possibilities need to be distinguished. If $(\TT^x, \TT^y) \propto (1, i)$, the stress tensor correspond to a dilation and/or a rotation (odd stress) depending on the phase. In this case it is independent of the orientation: $(1 , i)\mapsto (1, i)$.
    On the other hand if $(\TT^x,\TT^y) \propto (1,-i)$ the stress tensor has only pure shear components. It can be checked that it transforms as $(1, -i)\mapsto e^{2i\Theta}(1, -i)$.
    \item The force from Eq.~\eqref{smeq:fr_other} is anisotropic when $n\neq 1$. There is a specific angle (modulo $2\pi/(n-1)$) for which this force is radial: $\theta_0 = \alpha_n/(n-1)$. To consider a rotated system with a physically unchanged phase of the interactions, this angle needs to be rotated (in the direction opposite to the rotation of the particles). This leads to the mapping $\alpha_n \mapsto \alpha_n - (n-1)\Theta$ (valid even if $n=1$).
\end{enumerate} 

Finally, for a crystal of orientation $\Theta$, Eq.~\eqref{smeq:nodeform2} becomes
\begin{align}\label{smeq:nodeform3}
    \begin{pmatrix}
     \TT^x \\\TT^y
    \end{pmatrix}^0
    = \frac{\sqrt{3}f(a)e^{-i\alpha_n}}{a}
    \left\{\delta_{n, 6p+1} e^{i(n-1)\Theta} \begin{pmatrix}1 \\ i\end{pmatrix} 
    + \delta_{n, 6p+5}e^{i(n+1)\Theta} \begin{pmatrix}1 \\ -i\end{pmatrix}\right\},
\end{align}
in agreement with Eq.~\eqref{smeq:stress_nodef2} (withing a nearest-neighbors approximation). \\

\noindent \textbf{Dilation. } A dilation of magnitude $D_\delta$ corresponds to multiplying all lengths by a factor $(1+D_\delta)$. 
The final result is readily obtained from Eq.~\eqref{smeq:nodeform3} by making the substitution $a\mapsto a(1+D_\delta)$ at linear order in $D_\delta$:
\begin{equation}
    \begin{pmatrix}
     \delta\TT^x \\ \delta\TT^y
    \end{pmatrix}^\mathrm{dilat}
    = D_\delta \frac{\sqrt{3}e^{-i\alpha_n}}{a}\left[af'(a) - f(a)\right]
    \left\{\delta_{n, 6p+1} e^{i(n-1)\Theta} \begin{pmatrix}1 \\ i\end{pmatrix} 
    + \delta_{n, 6p+5}e^{i(n+1)\Theta} \begin{pmatrix}1 \\ -i\end{pmatrix}\right\}.
\end{equation}
This result is identical to Eq.~\eqref{smeq:stress_dilat2} (for nearest-neighbor interactions). \\

\noindent\textbf{Rotation. }  A rotation of angle $D_\omega$ corresponds to the substitution $\Theta \mapsto \Theta + D_\omega$ at linear order in $D_\omega$ in Eq.~\eqref{smeq:nodeform3}.
We obtain
\begin{align}
    \begin{pmatrix}
     \delta\TT^x \\ \delta\TT^y
    \end{pmatrix}^\mathrm{rot}
    = D_\omega\frac{\sqrt{3}f(a) i e^{-i\alpha_n}}{a}
    \left\{(n-1)\delta_{n, 6p+1} e^{i(n-1)\Theta} \begin{pmatrix}1 \\ i\end{pmatrix} 
    + (n+1) \delta_{n, 6p+5}e^{i(n+1)\Theta} \begin{pmatrix}1 \\ -i\end{pmatrix}\right\},
\end{align}
which is the same result as Eq.~\eqref{smeq:stress_rot2} (for nearest-neighbors). \\

\noindent\textbf{Shear deformations. } A shear deformation of amplitude $D_s$ with shear angle $\gamma$ corresponds to the complex map $z\mapsto z' = z + D_s e^{2i\gamma} z^\ast$ (where $z^\ast$ is the complex conjugate of $z$).
The force from Eq.~\eqref{smeq:fr_other} is modified as
\begin{equation}\label{smeq:dif_force_shear}
    \FF(z') - \FF(z=re^{i\theta}) = \frac{D_s}{2} e^{i(n\theta-\alpha_n)}\left\{\left[f'(r) + nf(r)\right] e^{2i(\gamma-\theta)} + \left[f'(r) - nf(r)\right] e^{-2i(\gamma-\theta)} \right\}.
\end{equation}
We study separately the first shear mode $\gamma=0$ and the second shear mode $\gamma=\pi/4$.

In the case $\gamma=0$, the two cuts that we consider  are still normal to $\hat{\vec x}$ and $\hat{\vec y}$. But one can check that the unit lengths associated with them are modified asymmetrically: $\ell_x \mapsto \ell_x(1-D_s)$ and $\ell_y \mapsto \ell_y(1+D_s)$. Using Eqs.~\eqref{smeq:TxTy} and~\eqref{smeq:dif_force_shear}, we obtain
\begin{multline}
    \begin{pmatrix}
     \delta\TT^x \\ \delta\TT^y
    \end{pmatrix}^\mathrm{shear1}
    = D_s \frac{\sqrt{3}e^{-i\alpha_n}}{2a}
    \bigg\{
    \left[af'(a) + nf(a)\right] \left[\delta_{n-2, 6p+1}  \begin{pmatrix}1 \\ i\end{pmatrix} 
    +\delta_{n-2, 6p+5}\begin{pmatrix}1 \\ -i\end{pmatrix}\right] \\
    + \left[af'(a) - nf(a)\right] \left[\delta_{n+2, 6p+1}  \begin{pmatrix}1 \\ i\end{pmatrix} 
    +\delta_{n+2, 6p+5}\begin{pmatrix}1 \\ -i\end{pmatrix}\right] 
    +2f(a) \left[\delta_{n, 6p+1}  \begin{pmatrix}1 \\ -i\end{pmatrix} 
    +\delta_{n, 6p+5}\begin{pmatrix}1 \\ i\end{pmatrix}\right]
    \bigg\}.
\end{multline}
which is Eq.~\eqref{smeq:stress_shear2} for $\gamma=0$, $\Theta=0$, and nearest-neighbors interactions only.

We now investigate the second shear mode $\gamma=\pi/4$. The horizontal and vertical lines of atoms are now rotated. 
Instead of considering cuts normal to $\hat{\vec x}=1$ and $\hat{\vec y}=i$, the new cuts are normal to $\hat{\vec x}' = e^{-iD_s}$ and $\hat{\vec y}' = ie^{iD_s}$.
Contrary to the first shear mode, the unit lengths $\ell_x$ and $\ell_y$ are unchanged. From Eqs.~\eqref{smeq:TxTy} and~\eqref{smeq:dif_force_shear}, we obtain
\begin{multline}
    \begin{pmatrix}
     \delta\TT^{x'} \\ \delta\TT^{y'}
    \end{pmatrix}^\mathrm{shear2}
    = D_s \frac{\sqrt{3}ie^{-i\alpha_n}}{2a}
    \bigg\{
    \left[af'(a) + nf(a)\right] \left[\delta_{n-2, 6p+1}  \begin{pmatrix}1 \\ i\end{pmatrix} 
    +\delta_{n-2, 6p+5}\begin{pmatrix}1 \\ -i\end{pmatrix}\right] \\
    - \left[af'(a) - nf(a)\right] \left[\delta_{n+2, 6p+1}  \begin{pmatrix}1 \\ i\end{pmatrix} 
    +\delta_{n+2, 6p+5}\begin{pmatrix}1 \\ -i\end{pmatrix}\right]
    \bigg\}.
\end{multline}
The forces in the $(x, y)$ coordinates are obtained by noting that at linear order $\hat{\vec x} = \hat{\vec x}' + D_s \hat{\vec y}'$ and $\hat{\vec y} = \hat{\vec y}' + D_s \hat{\vec x}'$. This leads to
\begin{align}
   \begin{pmatrix}
     \delta\TT^{x} \\ \delta\TT^{y}
    \end{pmatrix}^\mathrm{shear2} = \begin{pmatrix}
     \delta\TT^{x'} \\ \delta\TT^{y'}
    \end{pmatrix}^\mathrm{shear2} + D_s \begin{pmatrix}
     \TT^{y} \\ \TT^{x}
    \end{pmatrix}^0
\end{align}
which immediately gives Eq.~\eqref{smeq:stress_shear2} for $\gamma=\pi/4$, $\Theta=0$.

Finding the expression for an arbitrary orientation $\Theta$ is straightforward but tedious. 
One first needs to write the result for an arbitrary shear angle $\gamma$: $\delta\TT^{x/y} = \delta\TT^{x/y,\mathrm{shear1}}\cos(2\gamma) + \delta\TT^{x/y,\mathrm{shear2}}\sin(2\gamma)$.
Then the rotation of angle $\Theta$ needs to be done carefully: (i) the stress tensor is rotated according to the rules $(1, i)\mapsto (1,i)$ and $(1, -i)\mapsto e^{2i\Theta} (1, -i)$; (ii) the phase is rotated as $\alpha_n\mapsto \alpha_n - (n-1)\Theta$; (iii) the shear angle also needs to be rotated as $\gamma\mapsto \gamma-\Theta$. Once all of this is done, the full formula~\eqref{smeq:stress_shear2} is recovered in the nearest-neighbors appoximation. \\

In summary, we have presented  an alternative method to compute the stress-deformation constitutive relation of an hexagonal lattice made of particles interacting via parity-odd interactions. This method relies on the microscopic computation of the internal forces across two orthogonal surfaces. It allows us to recover the results Eqs.~\eqref{smeq:stress_nodef2}-\eqref{smeq:stress_shear2} without relying on the Irving-Kirkwood formula.

\subsection{Parity-even forces: body forces emerge from deformations} \label{ss:sym}
\subsubsection{Force-deformation relation}
We now consider the case of parity symmetric  forces $\FF\ind{S}(-\RR) = \FF\ind{S}(\RR)$.
We showed in Eq.~\eqref{smeq:stationarity} that these non-reciprocal forces may induce a net body force density in the crystal bulk.
Hence, when the system is deformed, at leading order in gradients, the constitutive relation defining the crystal mechanics relates a net body force force $\mathcal{F}$ to  deformations. At linear order, this relation reads
\begin{equation}
    \mathcal{F}
    = \mathcal{F}_0 
    + \begin{pmatrix}\ast & \ast & \ast & \ast \\ \ast & \ast & \ast & \ast\end{pmatrix}
    \begin{pmatrix}
    D_\delta \\ D_\omega \\ D_{s_1} \\ D_{s_2}
    \end{pmatrix},
\end{equation}
where $\mathcal{F}_0$ is the force supported by  an undeformed crystal and the $2\times 4$ matrix translates the linear relation between the local forces and  deformations. We  stress that this constitutive relation is intrinsically not invertible.

We focus again on the case of a hexagonal lattice (Fig.~\ref{fig:forces_stress}a), and  interactions of the form $\FF\ind{S}(r, \theta) = f(r)\Gg(\theta)$. In this case, the force $\mathcal{F}_0$ can be written using the same notations as in Eq.~\eqref{smeq:stress_nodef1}:
\begin{equation} \label{smeq:f_even}
    \mathcal{F}_0 = \sum_{\nu\in\mathbb{H}_{/6}} f(R^\nu) \sum_{k=0}^5 \Gg(\theta^{\nu,k})
\end{equation}
where the sum runs over the set $\mathbb{H}_{/6}$ of particles in the hexagonal lattice with radii $R^\nu$ and angles $\theta^\nu\in[0, \pi/3)$ (excluding the particle at the origin), and $\theta^{\nu,k} = \theta^\nu + k\pi/3$. 
Within a nearest neighbor approximation, this formula reduces to
\begin{equation}
    \mathcal{F}
 \overset{nn}{=} f(a) \sum_{k=0}^5 \Gg\left(\Theta + k\frac{\pi}{3}\right)
\end{equation}
where $a$ is the lattice spacing  and $\Theta$ its orientation.

Following the same procedure as in the previous section leading to  Eqs.~\eqref{smeq:stress_dil1}-\eqref{smeq:stress_shear1}, we can compute the change in the net force arising from deformations $\delta\mathcal{F}$ for  dilation, rotation and shear deformations. We find:
\begin{align}
\label{smeq:f_even_dilat}
    \delta\mathcal{F}^\mathrm{(dilat)} &= D_\delta \sum_{\nu\in\mathbb{H}_{/6}} R^\nu f'(R^\nu) \sum_{k=0}^5 \Gg(\theta^{\nu,k}) \\
    \label{smeq:f_even_rot}
    \delta\mathcal{F}^\mathrm{(rot)} &= D_\omega \sum_{\nu\in\mathbb{H}_{/6}} f(R^\nu) \sum_{k=0}^5 \Gg'(\theta^{\nu,k}) \\
\label{smeq:f_even_shear}
    \delta\mathcal{F}^\mathrm{(shear)} &= D_s \sum_{\nu\in\mathbb{H}_{/6}} \left[ 
    R^\nu f'(R^\nu) \sum_{k=0}^5 \Gg(\theta^{\nu,k}) \cos\left[2(\gamma-\theta^{\nu,k})\right]
    + f(R^\nu) \sum_{k=0}^5 \Gg'(\theta^{\nu,k}) \sin\left[2(\gamma-\theta^{\nu,k})\right]
    \right].
\end{align}
To gain some physical insight into these atypical constitutive relations, we first investigate in details the case of dipolar interactions (angular mode $n=2$) before addressing the general case of  angular modes of arbitrary (even) symmetry.

\subsubsection{Dipolar symmetry}
Let us consider a a dipolar force ($n=2$). In this simple case, we have
\begin{align}
    \label{smeq:dipole}
    \Gg(\theta) = e^{i(2\theta - \alpha_2)}
\end{align}
where $\alpha_2$ is the orientation of the dipole. It typically corresponds to the hydrodynamic interactions experienced by colloidal particles driven in a shallow Hele-Shaw channel~\cite{Desreumaux_2012}. From Eq.~\eqref{smeq:f_even} we quickly see that in the absence of deformations, the symmetry of the hexagonal lattice implies that no net force exists: $\mathcal{F} = 0$.
Similarly, from Eqs.~\eqref{smeq:f_even_dilat}-\eqref{smeq:f_even_rot}, there is also no force associated with dilation and rotation. The system  only responds to shear deformations.
Using Eqs.~\eqref{smeq:f_even_shear} and~\eqref{smeq:dipole}, we find
\begin{align} \label{smeq:f_dipole}
    \mathcal{F}^{(2)} & = \delta\mathcal{F}^\mathrm{(shear)} = D_s K_2 e^{i(2\gamma-\alpha_2)}, \\
    K_2 &= 3 \sum_{\nu\in\mathbb{H}_{/6}} \left[R^\nu f'(R^\nu) + 2 f(R^\nu)\right].
\end{align}
For nearest neighbor interactions only, $K_2 \overset{nn}{=} 3[a f'(a) + 2f(a)]$.
We use here the $(2)$ label to emphasize the dependence of the body force on the order $n$ of the angular symmetry. 
We also note that the $\mathcal{F}^{(2)}$ is   independent of the orientation $\Theta$ of the crystal.
The case of the drag force induced by 2D dipolar flows (potential flows)  corresponds to $f(r) = A r^{-2}$. 
We check numerically that in this case the sum converges to a finite value so that the relation Eq.~\eqref{smeq:f_dipole} remains valid, with $K_2 \approx -3.6 A/a^2$.

Finally, our results can be summarized by the following linear relation between the strain and the body-force acting on the crystal emerging from the microscopic violation of Newton's third law,
\begin{equation} \label{smeq:force_dipole}
    \mathcal{F}^{(2)}
    = K_2 \begin{pmatrix} 0 & 0 & \cos\alpha_2 & \sin\alpha_2 \\ 0 & 0 &-\sin\alpha_2 &  \cos\alpha_2 \end{pmatrix}
    \begin{pmatrix}
    D_\delta \\ D_\omega \\ D_{s_1} \\ D_{s_2}
    \end{pmatrix}.
\end{equation}

\subsubsection{Parity-even force of arbitrary angular symmetries}
We now consider generic parity-even forces given by
    $\Gg(\theta) = e^{i(n\theta - \alpha_n)}$,
with $n\in2\mathbb Z$.\\

\noindent{\bf Body force without deformations.} 
We first compute the force $\mathcal{F}_0$ acting on the crystal in the absence of deformation, using Eq.~\eqref{smeq:f_even},
\begin{equation}\label{smeq:f_nodef}
    \mathcal{F}_0^{(n)} = 6 \delta_{n, 6p} e^{-i\alpha_n} \sum_{\nu\in\mathbb{H}_{/6}} f(R^\nu) e^{in\theta^\nu}.
\end{equation}
There is a constant force without deformation if the inter-particle forces have the same angular symmetries as the lattice (sixfold symmetry).\\

\noindent{\bf Body force produced by dilations and rotations.} 
From Eqs.~\eqref{smeq:f_even_dilat} and \eqref{smeq:f_even_rot}, the change in net force due to a dilation (of amplitude $D_\delta)$ or a rotation (of amplitude $D_\omega$) is also non-zero only if $n=6p$:
\begin{align}
    \delta\mathcal{F}_\mathrm{dilat}^{(n)} &= 6 D_\delta\, \delta_{n, 6p} e^{-i\alpha_n} \sum_{\nu\in\mathbb{H}_{/6}} R^\nu f'(R^\nu) e^{in\theta^\nu}, \\
    \delta\mathcal{F}_\mathrm{rot}^{(n)} &= 6 D_\omega\, n \delta_{n, 6p} i e^{-i\alpha_n} \sum_{\nu\in\mathbb{H}_{/6}} f(R^\nu) e^{in\theta^\nu}.
\end{align}

\noindent{\bf Body force induced by shear deformations.} 
We then study the response to a shear deformation of amplitude $D_s$ and angle $\gamma$ from Eq.~\eqref{smeq:f_even_shear} and find,
\begin{multline}
    \delta\mathcal{F}_\mathrm{shear}^{(n)} = 3 D_s e^{-i\alpha_n} \bigg\{
    \delta_{n-2, 6p} e^{2i\gamma} \sum_{\nu\in\mathbb{H}_{/6}} \left[R^\nu f'(R^\nu) + n f(R^\nu) \right] e^{i(n-2)\theta^\nu} \\
    + \delta_{n+2, 6p} e^{-2i\gamma} \sum_{\nu\in\mathbb{H}_{/6}} \left[R^\nu f'(R^\nu) - n f(R^\nu) \right] e^{i(n+2)\theta^\nu}
    \bigg\}.
\end{multline}
Interactions corresponding to $n=6p+2$ or $n=6p+4$ result in response to shear deformations while those with $n=6p$ induce a net response to rotation and dilation.\\

The atypical mechanical response of driven crystal whose the constituents experience parity-even forces is summarized in Fig. 2 in the main text.
We recall the three possibles classes of constitutive relations below, depending on the angular symmetries of the non-reciprocal interactions:
\begin{align}
    \label{smeq:force_n0_0} 
    \mathcal{F}^{(n=6p)}  &= \begin{pmatrix} c_1 \\ c_2 \end{pmatrix}
    + \begin{pmatrix}c_1'  & -nc_2 & 0 & 0 \\
    c_2' & nc_1 & 0 & 0 \end{pmatrix}
    \begin{pmatrix}
    D_\delta \\ D_\omega \\ D_{s_1} \\ D_{s_2}
    \end{pmatrix} \\
    \label{smeq:force_n2_0}
    \mathcal{F}^{(n=6p+2)} &= \begin{pmatrix} 0 & 0 & a_1 & -a_2  \\ 0 & 0 & a_2 & a_1 \end{pmatrix}
    \begin{pmatrix}
    D_\delta \\ D_\omega \\ D_{s_1} \\ D_{s_2}
    \end{pmatrix}, \\
    \mathcal{F}^{(n=6p+4)} &= \begin{pmatrix} 0 & 0 & b_1 & b_2  \\ 0 & 0 & b_2 &  -b_1 \end{pmatrix}
    \begin{pmatrix}
    D_\delta \\ D_\omega \\ D_{s_1} \\ D_{s_2}
    \end{pmatrix},
    \label{smeq:force_n4_0}.
\end{align}
For nearest-neighbor interactions, the dependence in the phase $\alpha_n$ and the orientation $\Theta$ can be made explicit:
\begin{align} \label{smeq:force_n0_0_nn}
    \mathcal{F}^{(n=6p)} &\overset{nn}{=} K_n \begin{pmatrix} \cos(n\Theta-\alpha_n) \\ \sin(n\Theta-\alpha_n) \end{pmatrix}
    + \begin{pmatrix}K_n' \cos(n\Theta-\alpha_n)  & -nK_n\sin(n\Theta-\alpha_n) & 0 & 0 \\
    K_n' \sin(n\Theta-\alpha_n)   & nK_n \cos(n\Theta-\alpha_n) & 0 & 0 \end{pmatrix}
    \begin{pmatrix}
    D_\delta \\ D_\omega \\ D_{s_1} \\ D_{s_2}
    \end{pmatrix}  
    \\
    \label{smeq:force_n2_0_nn}
    \mathcal{F}^{(n=6p+2)} &\overset{nn}{=} K_n \begin{pmatrix} 0 & 0 & \cos[(n-2)\Theta - \alpha_n] & -\sin[(n-2)\Theta - \alpha_n]  \\ 0 & 0 & \sin[(n-2)\Theta - \alpha_n] &  \cos[(n-2)\Theta - \alpha_n] \end{pmatrix}
    \begin{pmatrix}
    D_\delta \\ D_\omega \\ D_{s_1} \\ D_{s_2}
    \end{pmatrix}, 
    \\
    \mathcal{F}^{(n=6p+4)} &\overset{nn}{=} K_n \begin{pmatrix} 0 & 0 & \cos[(n+2)\Theta - \alpha_n] & \sin[(n+2)\Theta - \alpha_n]  \\ 0 & 0 & \sin[(n+2)\Theta - \alpha_n] &  -\cos[(n+2)\Theta - \alpha_n] \end{pmatrix}
    \begin{pmatrix}
    D_\delta \\ D_\omega \\ D_{s_1} \\ D_{s_2}
    \end{pmatrix},
    \label{smeq:force_n4_0_nn}
\end{align}
with $K_{(n=6p)} = 6f(a)$, $K'_{(n=6p)} = 6af'(a)$, $K_{(n=6p+2)} = 3[af'(a) + n f(a)]$ and $K_{(n=6p+4)} = 3[af'(a) - n f(a)]$.

We notice that for $|n| > 2$, the response depends on the orientation $\Theta$ of the lattice.

\section{Self-propelled dislocations}

\subsection{Dislocation and Peach-Koehler force}
A dislocation in the plane is characterized by its Burgers vector $\bb$ defined as 
\begin{equation}
    b_j = \oint \partial_i u_j dr_i
\end{equation}
where the integration is performed on a closed loop in the counter-clockwise direction around the dislocation~\cite{Landau,Oswald,Braverman_2020}.
When a dislocation is subjected to an external stress $\sigma^\text{ext}$, elastic energy can be released by translating the dislocation.
The displacement $\delta\rr$ of the dislocation core and  the associated energy $\delta W$ are related by $\delta W = \FF^{PK}\cdot\delta\rr$, where $\FF^{PK}$ is the so-called Peach-Koehler force~\cite{Landau,Oswald,Braverman_2020}:
\begin{equation} \label{smeq:PK}
    F^{\rm PK}_i = \epsilon_{ij} \sigma^\text{ext}_{kj} b_k
\end{equation}
with $\epsilon_{ij}$ the Levi-Civita symbol. 

\subsection{Parity-antisymmetric interactions: self-propulsion of dislocations due to internal stresses} \label{smss:disloc_asym}
We showed  that parity-antisymmetric interactions can induce stresses in the absence of deformations, see Eqs.~\eqref{smeq:stress_1}-\eqref{smeq:stress_5}. The stresses, which exist regardless of the presence of a dislocation, contribute to the 
 energetic reasoning leading to the Peach-Koehler formula Eq.~\eqref{smeq:PK}. The resulting Peach-Koehler forces take the following forms:
\begin{enumerate}
    \item \textbf{If n=6p+1}, Eq.~\eqref{smeq:stress_1} predicts a stress tensor with a pressure term $P_\delta$ and a torque density $T_\omega$.
The orientation of the crystal is set by the Burgers vector $\Theta=\mathrm{angle}(\bb)$.
Using the Peach-Koehler formula~\eqref{smeq:PK} we obtain 
\begin{equation}
    \FF^{\rm PK} = -T_\omega\bb + P_\delta\bb^\perp
\end{equation}
with $b^\perp_i = \epsilon_{ij} b_j$.
The torque density leads to a Peach-Koehler force along the Burgers vector which  powers its gliding motion in agreement with our numerical simulations (see main text). The additional pressure term results in an effective climb force transverse to $\bf b$. The climb motion of a dislocation is however classically forbidden as it would involve a macroscopic displacement of all the atoms forming the lattice. As in equilibrium crystals the climb dynamics is therefore kinetically forbidden see e.g. Refs~\cite{Landau,Oswald}.

From Eq.~\eqref{smeq:stress_deform_iso}, we have $T_\omega = -\tilde P \sin \alpha_1$ (with $\tilde P = \sqrt{3}f(a)/a$ in the nearest-neighbors approximation). Therefore, the glide force is
\begin{equation}
    F^{\rm PK}_\mathrm{glide} = \tilde P\sin\alpha_1 b
\end{equation}
in agreement with the numerical results of Fig.~3c of the main text.

\item 
\textbf{If n=6p+3}, Eq.~\eqref{smeq:stress_3} predicts an absence of net stress in the undeformed system. The Peach-Koehler force vanishes yet we do observe dislocations gliding in our simulations (see main text). To account  for this dynamics we  extend the simple Peach-Koehler picture in subsection~\ref{smss:quad}. 

\item
\textbf{If n=6p+5}, Eq.~\eqref{smeq:stress_5} gives a stress tensor $\sigma^{\rm ext} = S(c_5 \tau_{s_1} + s_5 \tau_{s_2})$ with $c_5 = \cos(6\Theta-\alpha_5)$ and $s_5 = \sin(6\Theta-\alpha_5)$
and $S = \sqrt{3} f(a)/a$ (in the nearest-neighbors approximation).
The orientation of the crystal is set by the Burgers vector $\Theta=\mathrm{angle}(\bb)$.
Eq.~\eqref{smeq:PK} gives the following components for the Peach-Koehler force:
\begin{align}
    F^{\rm PK}_1 &= S s_n b_1 - S c_n b_2, &
    F^{\rm PK}_2 &= -S c_n b_1 - S s_n b_2.
\end{align}
To gain more insight into these formula, it is worth considering the lowest order mode $n=5$ with an horizontal Burgers vector, which yields
\begin{align}
    F^{\rm PK}_1(n=5, \bb = b\hat\ee_1) &= S b \sin\alpha_5, &
    F^{\rm PK}_2(n=5, \bb = b\hat\ee_1) &= S b \cos\alpha_5.
\end{align}
The glide component $F^{\rm PK}_1$ is proportional to the sine of the phase $\alpha_5$ in excellent agreement with our numerical simulations (see main text).
\end{enumerate}

\subsection{Parity-symmetric interactions: self-generated force field}
We now turn to parity-symmetric interactions and explain the motion of the dislocations in two steps. We first consider a dislocation in an elastic medium and compute the force field generated by parity-symmetric interactions.
We then explain how this force field leads to a glide of the dislocation.

\subsubsection{Force field generated by a dislocation}
Let us consider a dislocation of Burgers vector $\bb = b\hat\ee_1$ stabilized by shear and bulk elasticity. The displacement field reads~\cite{Landau,Oswald,Braverman_2020}
\begin{align}
    u_1(r, \theta) &= \frac{b}{2\pi}\left(\theta + \frac{1+\nu}{2}\cos\theta\sin\theta  \right), \\
    u_2(r, \theta) &= -\frac{b}{2\pi}\left[\frac{1-\nu}{2} \ln r + \frac{1+\nu}{2}\cos^2\theta\right],
\end{align}
where $\nu = (B-\mu)/(B+\mu)$ is the Poisson ratio.
The deformation tensor (Eq.~\eqref{smeq:deform_compos}) is then given by
\begin{equation} \label{smeq:disloc_deform_}
    \begin{pmatrix}
    D_\delta \\ D_\omega \\ D_{s_1} \\ D_{s_2}
    \end{pmatrix}(r, \theta)
    = \frac{b}{4\pi r}
    \begin{pmatrix}
    -(1-\nu)\sin\theta \\ 2\cos\theta  \\
    -(1+\nu)\cos\theta\sin(2\theta) \\ (1+\nu)\cos\theta\cos(2\theta)
    \end{pmatrix}.
\end{equation}

\begin{figure}
    \centering
    \includegraphics[scale=1]{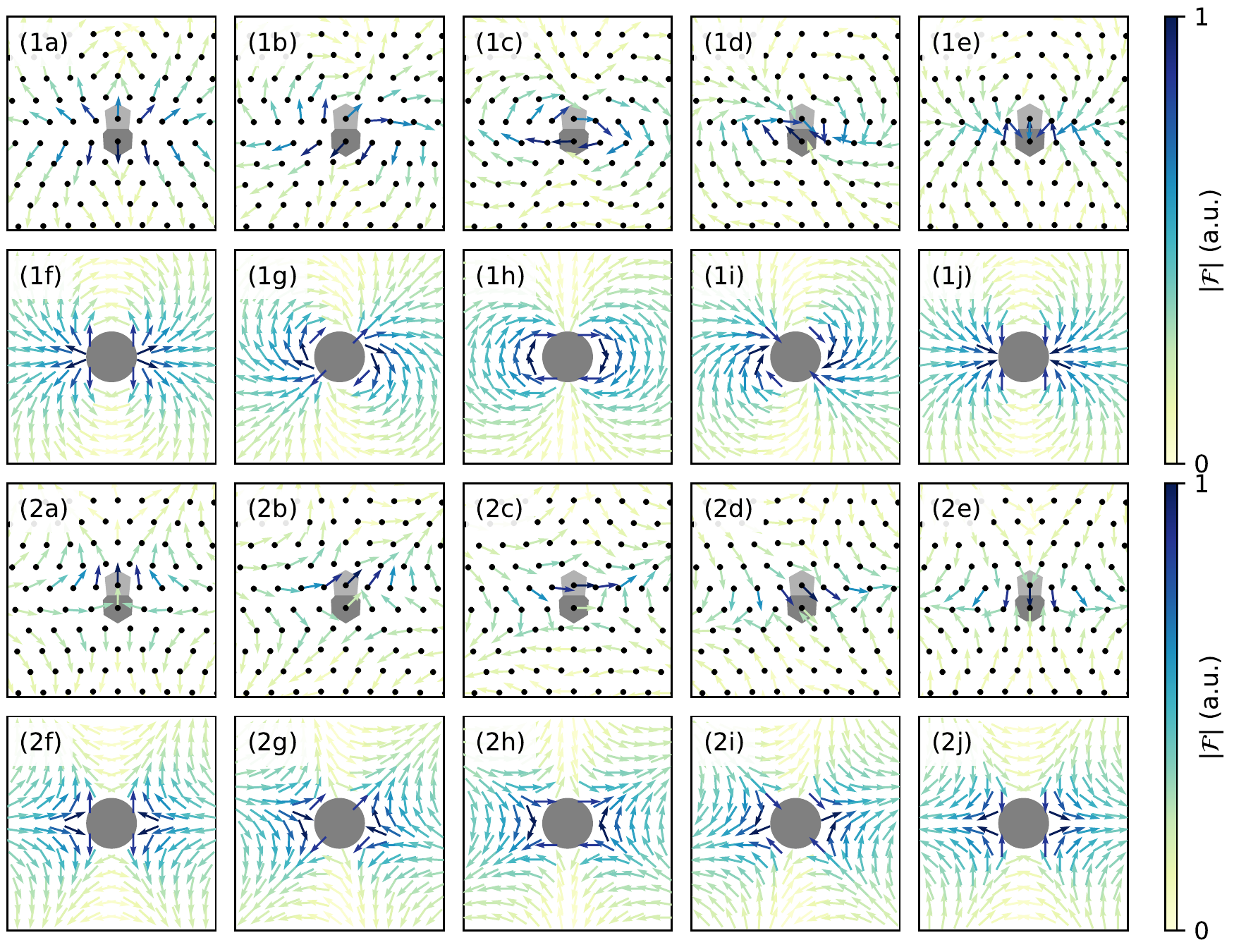}
    \caption{Force field induced by parity-symmetric interactions around a dislocation.
    Particles with 5 and 7 neighbors are shown in respectively in light gray and dark grey.
    (1a)-(1e) Dipolar symmetry ($n=2$), $\FF(r, \theta) = e^{2i(\theta-\alpha_2)}/r^2$. From (1a) to (1e) $\alpha_2 = -\pi/2, -\pi/4, 0, \pi/4, \pi/2$. 
    (1f)-(1j) Corresponding theoretical prediction given by Eq.~\eqref{smeq:force_n2}.
    (2a)-(2e) Hexapolar symmetry ($n=4$), $\FF(r, \theta) = e^{4i(\theta-\alpha_4)}/r^4$ with
    $\alpha_4 = -\pi/2, -\pi/4, 0, \pi/4, \pi/2$.
    (2f)-(2j) Corresponding prediction according to Eq.~\eqref{smeq:force_n4}.
    Note that the coefficients $K_2$ and $K_4$ are both negative.
    }
    \label{fig:disloc}
\end{figure}

We now address the effect of (small) parity-symmetric interactions $\FF(\rr) = f_n(r) e^{i(n\theta - \alpha_n)}$ with $n$ even. Eqs.~\eqref{smeq:force_n0_0_nn}-\eqref{smeq:force_n4_0_nn} (nearest-neighbor interactions) along with
Eq.~\eqref{smeq:disloc_deform_} allow us to compute the force field $\mathcal{F}(r,\theta)$ generated by the elastic deformations induced by the dislocation.
We note that the orientation of the crystal is set by the Burgers vector ($\Theta = 0$).
We can again distinguish three cases depending on the orientational symmetry of the non-reciprocal forces:
\begin{align}
    \label{smeq:force_n0}
    \mathcal{F}^{(n=6p)}\ind{disloc}(r, \theta) &= \left[K_n - \frac{b(1-\nu)\sin\theta}{4\pi r}K_n' \right] \begin{pmatrix}\cos\alpha_n \\ -\sin\alpha_n \end{pmatrix}
    + \frac{b\cos\theta}{2\pi r} n K_n \begin{pmatrix}\sin\alpha_n \\ \cos\alpha_n \end{pmatrix}, \\
    \label{smeq:force_n2}
    \mathcal{F}^{(n=6p+2)}\ind{disloc}(r, \theta) &= K_n \frac{b (1+\nu)}{4\pi} \frac{\cos\theta}{r}
    \begin{pmatrix}
    -\sin(2\theta-\alpha_n) \\ \cos(2\theta-\alpha_n)
    \end{pmatrix}, \\ 
    \label{smeq:force_n4}
    \mathcal{F}^{(n=6p+4)}\ind{disloc}(r, \theta) &= -K_n \frac{b (1+\nu)}{4\pi} \frac{\cos\theta}{r}
    \begin{pmatrix}
    \sin(2\theta+\alpha_n) \\ \cos(2\theta+\alpha_n)
    \end{pmatrix}.
\end{align}
In Fig.~\ref{fig:disloc}, we compare this theoretical prediction relying on a continuous description of the non-equilibrium crystal with the microscopic forces acting on each particle in the crystal that are computed numerically  for $n=2$ and $4$ (note that $K_2 < 0$ and $K_4 < 0$).

\subsubsection{Force gradient around the dislocation and spontaneous motion}
Fig.~\ref{fig:disloc} reveals that the net force arising from the momentum exchange between the crystal particles and their surrounding matrix varies in space around the dislocation center. Focusing on the situation where $\alpha_2 = 0$ and $\alpha_4=0$ (third column), one can clearly notice a strong shear contribution to the  force-field gradient (gradient of horizontal force in the vertical direction). The Peach-Koehler formula tells us that a local shear  promotes the gliding motion of an isolated dislocation, as confirmed by direct experimental observations in colloidal crystals~\cite{Irvine_2013}. 

The force field acting on the crystal defined by Eqs.~\eqref{smeq:force_n0}-\eqref{smeq:force_n4}  diverges when $r\to 0$. To define a force gradient at the origin we therefore need to use a regularization scheme. We define the regularized force gradient at the origin as an integral over a circle of radius
$r_0 \simeq b$, 
\begin{equation} \label{smeq:regul}
    \left. \frac{\partial\mathcal{F}_i}{\partial x_j}\right|_\mathrm{reg} 
    = \frac{1}{2\pi b} \int_0^{2\pi} \mathcal{F}_i(b, \theta) \hat e_j(\theta) d\theta
\end{equation}
where $\hat \ee(\theta) = (\cos\theta, \sin\theta)$.
We stress that the  regularization radius $r_0$ is an arbitrary cut-off scale. Physically $r_0$ represents the distance at which the far-field expression (that diverges when $r\to 0$) ceases to be valid. Its scale is thus set by the amplitude of the Burgers vector, we hence set $r_0\simeq b$.
In principle, the prediction of the gliding speed of a dislocation would require computing the dislocation mobility coefficient  using a consistent choice of  cut-off scale. 
Computing the dislocation mobility is a very challenging task which goes beyond the scope of our letter. 
In all that follows our  predictions concern the angular variations of the gliding speed, not its absolute magnitude.

The force gradients effectively define an  external stress tensor $\sigma_{ij}=\partial\mathcal{F}_i/\partial x_j$ acting around the dislocation core. Using  Eqs.~\eqref{smeq:force_n0}-\eqref{smeq:force_n4}, we find
\begin{align}
    \left. \frac{\partial\mathcal{F}_i}{\partial x_j}\right|_\mathrm{reg}^{(n=6p)}
    &= \frac{K_n'(1-\nu)}{8\pi} \begin{pmatrix}
     0 & -\cos\alpha_n \\ 0 & \sin\alpha_n
    \end{pmatrix}
    + \frac{nK_n}{4\pi} \begin{pmatrix}
     \sin\alpha_n & 0 \\ \cos\alpha_n & 0
    \end{pmatrix}, \\
    \left. \frac{\partial\mathcal{F}_i}{\partial x_j}\right|_\mathrm{reg}^{(n=6p+2)}
    &= \frac{K_n(1+\nu)}{16\pi} \begin{pmatrix}
     \sin\alpha_n & -\cos\alpha_n \\ \cos\alpha_n & \sin\alpha_n
    \end{pmatrix}, \\
    \left. \frac{\partial\mathcal{F}_i}{\partial x_j}\right|_\mathrm{reg}^{(n=6p+4)}
    &= -\frac{K_n(1+\nu)}{16\pi} \begin{pmatrix}
     \sin\alpha_n & \cos\alpha_n \\ \cos\alpha_n & -\sin\alpha_n
    \end{pmatrix}.
\end{align}

According to the Peach-Koehler formula~\eqref{smeq:PK}, the component $\partial\mathcal{F}_1/\partial x_2$ produces a glide force and  $\partial\mathcal{F}_1/\partial x_1$ a climb force (for a dislocation having Burgers vector pointing along $x_1$). We therefore define a force  akin to an equilibrium Peach-Koehler force, $\FF^\mathrm{PK}$ acting on the dislocation as
\begin{equation}
    F^\mathrm{PK}_i = \epsilon_{ij} \left. \frac{\partial\mathcal{F}_k}{\partial x_j}\right|_\mathrm{reg} b_k.
\end{equation}

For a Burgers vector $\bb = b \hat\ee_1$, we find
\begin{align}
\label{smeq:Fpk_0}
    \FF^{(n=6p)}\ind{PK} &= -\frac{K_n'(1-\nu)b}{8\pi}
    \begin{pmatrix} \cos\alpha_n \\ 0 \end{pmatrix} 
    + \frac{n K_n b}{4\pi}
    \begin{pmatrix} 0 \\ \sin\alpha_n \end{pmatrix}, \\
     \FF^{(n=6p+2)}\ind{PK} &= -\frac{K_n(1+\nu)b}{16\pi} \begin{pmatrix}
    \cos\alpha_n \\ \sin\alpha_n 
    \end{pmatrix}, \\
\label{smeq:Fpk_4}
    \FF^{(n=6p+4)}\ind{PK} &= -\frac{K_n(1+\nu)b}{16\pi} \begin{pmatrix}
    \cos\alpha_n \\ -\sin\alpha_n
    \end{pmatrix}.
\end{align}
In all the cases, the glide force is proportional to $\cos\alpha_n$, which is again in excellent agreement with our numerical measurements of the dislocation speeds reported in the main text.

\subsection{Specific case of interactions having a quadrupolar symmetry} \label{smss:quad}
In our numerical simulations, we observe the glide of  dislocations deforming elastic media perturbed by microscopic non-reciprocal forces  having quadrupolar symmetry ($n=3$). 
Our numerical simulations indicate that the dislocation speed scales with $\sin\alpha_3$, where $\alpha_3$ is the phase angle of the quadrupolar force. 
In the absence of internal odd and shear stresses, this observation cannot be explained within the Peach-Koehler framework detailed in subsection~\ref{smss:disloc_asym}.
 To account for our quantitative observation, we compute the stress tensor arising from the non-reciprocal interactions only. From Eq.~\eqref{smeq:stress_3} and Eq.~\eqref{smeq:disloc_deform_} we find
\begin{equation}
    \begin{pmatrix}
    \sigma_\delta  \\ \sigma_\omega  \\ \sigma_{s_1} \\ \sigma_{s_2}
    \end{pmatrix}(r, \theta)
    = 
    \frac{b(1+\nu)\cos\theta}{4\pi r}
    \begin{pmatrix}
    -C\sin(2\theta-\alpha_3) \\ C\cos(2\theta-\alpha_3)  \\ \tilde\mu \sin(6\Theta-2\theta-\alpha_3) \\ -\tilde\mu \cos(6\Theta-2\theta-\alpha_3)
    \end{pmatrix}
\end{equation}
where we used the nearest-neighbors expression and we defined $C=\sqrt{3}[af'(a)+3f(a)]/(2a)$ and $\tilde\mu = \sqrt{3}[af'(a)-3f(a)]/(2a)$.
We now arbitrarily divide the system into two parts: the core of the dislocation corresponding to $r<b$, and the outer region corresponding to  $r>b$. By definition of the stress tensor, the force that the outer region  applies on the dislocation core at a point $b\hat\ee(\theta)$ is given by $\TT(b, \theta) = \sigma(b,\theta)\cdot \hat\ee(\theta)$ where $\hat\ee(\theta) = (\cos\theta, \sin\theta)$. Its expression is
\begin{equation} \label{smeq:T3}
    \TT(b, \theta) = \frac{(1+\nu)\cos\theta}{4\pi} \begin{pmatrix}
     -C\sin(3\theta-\alpha_3) + \tilde\mu \sin(6\Theta-3\theta-\alpha_3) \\
     C\cos(3\theta-\alpha_3) - \tilde\mu \cos(6\Theta-3\theta-\alpha_3)
    \end{pmatrix}.
\end{equation}
We conclude that there  exists a nonvanishing force distribution $\TT(b, \theta)$ acting  on the dislocation  core. This begs for two questions: Does it result in a net force? Does this force distribution induce a net stress acting on the core region? At first sight, both questions seems to have a negative answer. If the crystal orientation $\Theta$ is taken to be a constant, then  
$\int_0^{2\pi} \TT(b, \theta) d\theta$ and the force gradient $\left. \frac{\partial T_i}{\partial x_j}\right|_\mathrm{reg}$ defined from Eq.~\eqref{smeq:regul} both vanish.

The key point is that the orientation of the crystal around a dislocation is not constant. From Eq.~\eqref{smeq:disloc_deform_}, we know that the rotation $D_\omega$ is non zero. The orientation of the crystal around the core of the dislocation is given by
\begin{equation} \label{smeq:theta3}
    \Theta(b, \theta) = D_\omega(b, \theta) = \omega\cos\theta
\end{equation}
with $\omega = 1/(2\pi)$.

Combining Eqs.~\eqref{smeq:T3} and \eqref{smeq:theta3}, we then find that the integrated force does not vanish anymore and results in  elastic deformations.
More importantly, combining Eqs.~\eqref{smeq:T3} and \eqref{smeq:theta3} we find a non zero force gradient, viz. a stress, having a non-vanishing simple shear component
\begin{equation}
    \left. \frac{\partial T_1}{\partial x_2}\right|_\mathrm{reg} = \frac{1}{2\pi b} \int_0^{2\pi} T_1(b,\theta) \sin\theta d\theta = S \sin\alpha_3
\end{equation}
with 
$S = \frac{(1+\nu)\tilde\mu}{4\pi b} \frac{1}{6\omega^2}\left[2J_3(6\omega) - 3\omega J_2(6\omega)\right]$ where $J_n$ is a Bessel function of the first kind.

Simply put, we find that the variation of the orientation of an elastic crystal perturbed by quadrupolar non-reciprocal interactions gives rise to a simple shear stress acting on the core of a dislocation. In turn, this shear powers the glide of isolated dislocations. Remarkably, the scaling of the driving stress with  $\sin\alpha_3$ agrees with the numerical measurement of the dislocation speed.

\section{Numerical methods}
We consider a periodic box of size $(L_x, L_y) = (91.8, 52.8)$ including $N = 5582$ particles at positions $\RR_i$. Two dislocations of opposite Burgers vectors $\pm \hat \ee_1$ were generated at maximally distant positions $(0, 0)$ and $(L_x/2, L_y/2)$.
We use a forward Euler scheme to solve the overdamped dynamics
\begin{equation}
    \RR_i(t+\Delta t) = \RR_i(t) + \Delta t \sum_{i\neq j} \left\{ \FF\ind{repulsive}(\RR_i - \RR_j) + \FF\ind{non-rec}(\RR_i - \RR_j)\right\}
\end{equation}
where the repulsive forces come from a dipole-dipole potential $\phi(\rr) \sim r^{-3}$,
\begin{equation}
    \FF\ind{repulsive}(\rr) = \frac{A\ind{rep}}{\|\rr\|^4} \frac{\rr}{\|\rr\|}
\end{equation}
and the non-reciprocal forces are
\begin{equation}
    \FF\ind{non-rec}(r, \theta) = \frac{1}{r^n} \begin{pmatrix}
     \cos(n\theta - \alpha_n) \\ \sin(n\theta - \alpha_n)
    \end{pmatrix}
\end{equation}
for $n\geq 2$ and 
\begin{equation}
    \FF\ind{non-rec}(r, \theta) = e^{-\frac{r^2}{2}} \begin{pmatrix}
     \cos(n\theta - \alpha_1) \\ \sin(n\theta - \alpha_1)
    \end{pmatrix}
\end{equation}
for $n=1$.

In the case of long-range forces, we set a cutoff at a distance $r=L_y/2$, except in the case $n=2$ where we implemented an Ewald summation in the periodic box~\cite{FrenkelSmit}.
The strength of the repulsive forces is varied between $A\ind{rep}=1$ and $A\ind{rep}=50$ depending on the non-reciprocal forces (values given in the main text). The timestep is typically $\Delta t = 2\times 10^{-3}$, and the simulations were run over $2000$ iterations.
The dislocations were tracked by detecting the particles with $5$ and $7$ neighbors and their velocity is measured from a linear fit of their trajectory.

In the supplemental video, we use a smaller system for illustration purpose ($N=679$ particles in a box of size $(L_x, Ly)=(29.5, 20)$).
The phases of the non-reciprocal forces are respectively $\alpha_1 = \alpha_3 = \alpha_5 = \pi/2$, $\alpha_2 = 20 \pi/180$ and $\alpha_4 = \alpha_6 = 0$.
The radial dependence of the forces and the values of $A\ind{rep}$ are identical to the ones used in the main text (Fig. 3).